\documentclass[aps,prd,longbibliography,reprint,twocolumn,amsmath,amssymb,amsfonts,showpacs,footnote]{revtex4-1}

\usepackage{ifpdf}
\usepackage{dcolumn}
\usepackage{enumitem}
\usepackage{bm}
\usepackage{here}
\usepackage{float}

\floatstyle{plaintop}
\restylefloat{table}
\ifpdf

      \usepackage[pdftex]{graphicx}  
      \usepackage[pdftex]{epsfig}

      \usepackage[pdftex]{hyperref}
\hypersetup
	{
		colorlinks,%
		citecolor=blue,%
		linkcolor=red,%
		urlcolor=blue,%
	}

\else

      \usepackage[dvips]{graphicx}  
      \usepackage[dvips]{epsfig}

      \usepackage[dvipdfm]{hyperref}
      \hypersetup
	{
		colorlinks,%
		citecolor=blue,%
		linkcolor=red,%
		urlcolor=blue,%
	}

\fi

\begin{document}

\title{Electromagnetic radiation generated by a charged particle falling radially into a Schwarzschild black hole: A complex angular momentum description }

\author{Antoine Folacci}
\email{folacci@univ-corse.fr}
\affiliation{Equipe Physique
Th\'eorique, \\ SPE, UMR 6134 du CNRS
et de l'Universit\'e de Corse,\\
Universit\'e de Corse, Facult\'e des Sciences, BP 52, F-20250 Corte,
France}

\author{Mohamed \surname{Ould~El~Hadj}}
\email{med.ouldelhadj@gmail.com}
\affiliation{Equipe Physique
Th\'eorique, \\ SPE, UMR 6134 du CNRS
et de l'Universit\'e de Corse,\\
Universit\'e de Corse, Facult\'e des Sciences, BP 52, F-20250 Corte,
France}

\date{\today}

\begin{abstract}

By using complex angular momentum techniques, we study the electromagnetic radiation generated by a charged particle falling radially from infinity into a Schwarzschild black hole. We consider both the case of a particle initially at rest and that of a particle projected with a relativistic velocity and we construct complex angular momentum representations and Regge pole approximations of the partial wave expansions defining the Maxwell scalar $\phi_2$ and the energy spectrum $dE/d\omega$ observed at spatial infinity. We show, in particular, that Regge pole approximations involving only one Regge pole provide effective resummations of these partial wave expansions permitting us (i) to reproduce with very good agreement the black hole ringdown without requiring a starting time, (ii) to describe with rather good agreement the tail of the signal and sometimes the pre-ringdown phase, and (iii) to explain the oscillations in the electromagnetic energy spectrum radiated by the charged particle. The present work as well as a previous one concerning the gravitational radiation generated by a massive particle falling into a Schwarzschild black hole [A. Folacci and M. Ould El Hadj, Phys.\, Rev.\, D {\bf 98}, 064052 (2018), arXiv:1807.09056 [gr-qc]] highlight the benefits of studying radiation from black holes in the complex angular momentum framework (they obviously appear when the approximations obtained involve a small number of Regge poles and have a clear physical interpretation) but also to exhibit the limits of this approach (this is the case when it is necessary to take into account background integral contributions).

\end{abstract}

\maketitle

\tableofcontents

\section{Introduction}

In a recent article \cite{Folacci:2018sef}, we advocated for an alternative description of gravitational radiation from black holes (BHs) based on complex angular momentum (CAM) techniques, i.e., analytic continuation in the CAM plane of partial wave expansions, duality of the ${\cal S}$-matrix, effective resummations involving its Regge poles and the associated residues, Regge trajectories, semiclassical interpretations, etc. In this previous article as well as in more recent works \cite{Folacci:2019cmc,Folacci:2019vtt} where we have provided CAM and Regge pole analyses of scattering of scalar, electromagnetic and gravitational waves by a Schwarzschild BH, we have justified the interest of such an approach in the context of BH physics and we shall not return here on this subject. We refer the interested reader to these articles and, more particularly, to the introduction of Ref.~\cite{Folacci:2018sef} as well as to references therein for other works dealing with the CAM approach to BH physics.

In the present article, by using CAM techniques, we shall revisit the problem of the electromagnetic radiation generated by a charged particle falling radially into a Schwarzschild BH. We shall consider both the case of a particle initially at rest and that of a particle projected with a relativistic or an ultra-relativistic velocity and we shall construct CAM representations and Regge pole approximations of the partial wave expansions defining the Maxwell scalar $\phi_2$ and the energy spectrum $dE/d\omega$ observed at spatial infinity. This work extends our previous work concerning the CAM and Regge pole analyses of the gravitational radiation generated by a massive particle falling into a Schwarzschild BH \cite{Folacci:2018sef}. Both highlight the benefits of working within the CAM framework and strengthen our opinion concerning the interest of the Regge pole approach for describing radiation from BHs.

Problems dealing with the excitation of a BH by a charged particle and the generation of the associated electromagnetic radiation have been considered, since the early 1970s, in the literature (for pioneering works on this subject, see the lectures by Ruffini~\cite{Ruffini:1973pta} in Ref.~\cite{DeWitt:1973uma} and references therein, as well as Refs.~\cite{Ruffini:1972pw,Ruffini:1972uh,Tiomno:1972dq,Cardoso:2003cn} for articles directly relevant to our study) and, currently, an ever-increasing importance is given to them. Indeed, such problems are of great interest with the emergence of multimessenger astronomy which combines the detection and analysis of gravitational waves with those of other types of radiation for a better understanding of our ``violent Universe'' but also in order to test the BH hypothesis and Einstein's general relativity in the strong-field regime (see, e.g., Refs.~\cite{Psaltis2008,Johannsen_2012,Bambi:2015kza}). In this context, it is particularly interesting to study the electromagnetic partner of the gravitational radiation generated during the accretion of a charged fluid by a BH \cite{Degollado:2014dfa,Moreno:2016urq}.

Our paper is organized as follows. In Sec.~\ref{SecII}, we first construct the Maxwell scalar $\phi_2$ describing the outgoing electromagnetic radiation at infinity which is generated by a charged particle falling radially into a Schwarzschild BH. To do this, by using Green's function techniques, we solve in the frequency domain the Regge-Wheeler equation for arbitrary $(\ell,m)$ modes and we proceed to their regularization. We also extract from the multipole expansion of $\phi_2$ the quasinormal ringdown of the BH. In Sec.~\ref{SecIII}, we provide two different CAM representations of the multipolar waveform $\phi_2$: the first one is based on the Poisson summation formula~\cite{MorseFeshbach1953} while the second one is constructed from the Sommerfeld-Watson transformation~\cite{Watson18,Sommerfeld49,Newton:1982qc}. From each of them, we extract, as approximations of $\phi_2$, the Fourier transform of a sum over Regge poles and Regge-mode excitation factors. It is important to note that, in order to evaluate numerically these two Regge pole approximations, we need the Regge trajectories (i.e., the curves traced out in the CAM plane by the Regge poles and by the associated residues as a function of the frequency $\omega$). In Sec.~\ref{SecIV}, we numerically compare the multipolar waveform $\phi_2$ constructed by summing over a large number of partial modes (this is particularly necessary for a particle projected with a relativistic or an ultra-relativistic velocity) as well as the associated ringdown with the Regge pole approximations obtained in Sec.~\ref{SecIII}. This permits us to emphasize the benefits of working with these particular approximations of the Maxwell scalar $\phi_2$. In Sec.~\ref{SecV}, we focus on the electromagnetic energy spectrum $dE/d\omega$ radiated by the charged particle falling into the Schwarzschild BH and we numerically compare it with its CAM representation obtained from the Poisson summation formula. In the Conclusion, we summarize the main results obtained and briefly discuss some possible extensions of our approach.

Throughout this article, we adopt units such that $G = c = \epsilon_0 = \mu_0 = 1$, we use the geometrical conventions of Ref.~\cite{Misner:1974qy} and we perform the numerical calculations using {\it Mathematica} \cite{Mathematica}. We, furthermore, consider that the exterior of the Schwarzschild BH is defined by the line element $ds^2= -f(r) dt^2+ f(r)^{-1}dr^2+ r^2 d\theta^2 + r^2 \sin^2\theta d\varphi^2$ where $f(r)=1-2M/r$ and $M$ is the mass of the BH while $t \in ]-\infty, +\infty[$, $r \in ]2M,+\infty[$, $\theta \in [0,\pi]$ and $\varphi \in [0,2\pi]$ are the usual Schwarzschild coordinates.

\section{Maxwell scalar $\phi_2$ and associated quasinormal ringdown}
\label{SecII}

In this section, we shall construct the Maxwell scalar $\phi_2$ describing the outgoing radiation at infinity due to a charged particle falling radially from infinity into a Schwarzschild BH. Moreover, we shall extract from the multipole expansion of $\phi_2$ the associated ringdown waveform.

\subsection{Multipole expansion of the Maxwell scalar $\phi_2$}
\label{SecIIa}

We consider a charged particle (we denote by $m_0$ its mass and by $q$ its electric charge) falling radially into a Schwarzschild BH. The timelike geodesic followed by such a particle is defined by the coordinates $t_{p}(\tau)$, $r_{p}(\tau)$, $\theta_{p}(\tau)$ and $\varphi_{p}(\tau)$ where $\tau$ is its proper time. Without loss of generality, we can consider that this particle moves in the BH equatorial plane along the positive $x$ axis and in the negative direction, i.e., we assume that $\theta_{p}(\tau)=\pi/2$, $\varphi_{p}(\tau)=0$ and $dr_{p}(\tau)/{d\tau} <0$. The functions $t_{p}(\tau)$, $r_{p}(\tau)$ as well as the function $t_{p}(r)$ can be then obtained from the geodesic equations (see, e.g., Ref.~\cite{Chandrasekhar:1985kt})
\begin{subequations}
\label{geodesic_equations}
\begin{equation}
f(r_{p})\frac{dt_{p}}{d\tau}=\frac{\cal{E}}{m_0}, \label{geodesic_1}
\end{equation}
\noindent and
\begin{equation}
\label{geodesic_3}
\left(\frac{dr_{p}}{d\tau}\right)^{2} -\frac{2M}{r_{p}}
=\left(\frac{\cal{E}}{m_0}\right)^{2}-1.
\end{equation}
\end{subequations}
Here, $\cal{E}$ is the energy of the particle. It is a constant of motion which can be related to the velocity $v_\infty$ of the particle at infinity and to the associated Lorentz factor $\gamma$ by
\begin{equation}
\label{tildeEand gamma}
\frac{\cal{E}}{m_0}=\frac{1}{\sqrt{1-(v_\infty)^2}}=\gamma.
\end{equation}

The electromagnetic radiation generated by this particle can be described by using the gauge-invariant formalism introduced by Ruffini, Tiomno and Vishveshwara in Ref.~\cite{Ruffini:1972pw} (see also Refs.~\cite{Cunningham:1978zfa,Cunningham:1979px}) and by working in the framework of the Newman-Penrose formalism (see, e.g., Chap.~8 of Ref.~\cite{Alcubierre:1138167}). We shall therefore focus on the Maxwell scalar $\phi_2$ which can be expressed at spatial infinity as
\begin{equation}\label{phi2_def}
\phi_2 = \frac{1}{2\sqrt{2}}\{\left(E_\theta - i E_\varphi \right) + i \left(B_\theta - i B_\varphi \right)\}
\end{equation}
where $E_\theta$, $E_\varphi$, $B_\theta$ and $B_\varphi$ denote the components of the electromagnetic field  $({\bf E}, {\bf B})$ observed for $r \to \infty $. Here, it is important to note that we have defined $\phi_2$ with respect to the null basis $(l,n,m,m^\ast)$ which is normalized such that the only nonvanishing scalar products involving the vectors of the tetrad are $l^\mu n_\mu = -1$ and $m^\mu m^\ast_\mu=1$ and which is given by (our conventions slightly differ from those of Ref.~\cite{Alcubierre:1138167})
\begin{subequations}
\begin{eqnarray} \label{NullTetrad}
& & l^\mu=\left(\frac{1}{f(r)},1,0,0 \right), \\
& & n^\mu=\frac{1}{2}\left(1,-\frac{1}{f(r)},0,0\right), \\
& & m^\mu=\frac{1}{\sqrt{2} r}\left(0,0,1, \frac{i}{\sin \theta}\right), \\
& & {m^\ast}^\mu=\frac{1}{\sqrt{2} r}\left(0,0,1, \frac{-i}{\sin \theta}\right).
\end{eqnarray}
\end{subequations}
We recall that the radially infalling particle only excites the even (polar) electromagnetic modes of the Schwarzschild BH and that, in the usual orthonormalized basis $({\bf {\hat{e}}_r},{\bf {\hat{e}}_\theta},{\bf {\hat{e}}_\varphi})$ of the spherical coordinate system, the components of the electric field can be expressed in terms of the gauge-invariant master functions $\psi_{\ell m} (t,r)$ and expanded on the (even) vector spherical harmonics $Y_\theta^{\ell m}(\theta,\varphi)$ and $Y_\varphi^{\ell m}(\theta,\varphi)$ in the form~\cite{Folacci:2018vtf}
\begin{equation}\label{ChampE_even}
{\mathbf E}=\left| \begin{array}{l}
 E_r = 0 \\
 E_\theta = -\frac{1}{r} \sum\limits_{\ell =1}^{+\infty}\sum\limits_{m=-\ell}^{+\ell} \frac{1}{\ell(\ell+1)}\, \partial_r\psi_{\ell m} \, Y_\theta^{\ell m} \\
 E_\varphi = -\frac{1}{r \sin\theta} \sum\limits_{\ell =1}^{+\infty}\sum\limits_{m=-\ell}^{+\ell} \frac{1}{\ell(\ell+1)} \, \partial_r \psi_{\ell m} \, Y_\varphi^{\ell m}, \\
\end{array}
\right.
\end{equation}
while the magnetic field $\mathbf{B}$ can be obtained from the Maxwell-Faraday equation and its components expressed in terms of those of the electric field. Indeed, for $r \to +\infty$, we have $\partial_t\psi_{\ell m} = - \partial_r\psi_{\ell m}$ and we can write
\begin{equation}\label{Relations_ChampB_even_odd}
{\mathbf B}=\left| \begin{array}{l}
 B_r = 0 \\
 B_\theta = - E_\varphi  \\
 B_\varphi = + E_\theta. \\
\end{array}
\right.
\end{equation}
It should be noted that the vector spherical harmonics appearing in Eq.~(\ref{ChampE_even}) are given in terms of the standard scalar spherical harmonics $Y^{\ell m}(\theta,\varphi)$ by
\begin{equation}\label{HSV_even}
Y_\theta^{\ell m} = \frac{\partial}{\partial \theta} Y^{\ell m} \quad \text{and} \quad Y_\varphi^{\ell m}= \frac{\partial}{\partial \varphi} Y^{\ell m}
 \end{equation}
and satisfy the ``orthonormalization'' relation
\begin{eqnarray}\label{HSV_even_Norm}
& & \int _{{\cal S}^2} d\Omega_2 ~\left[ Y_\theta^{\ell m}(\theta,\varphi) [Y_\theta^{\ell'
m'}(\theta,\varphi)]^* \quad \phantom{\frac{1}{\sin^2 \theta}} \right. \nonumber \\
& & \qquad \left. + \frac{1}{\sin^2 \theta} Y_\varphi^{\ell m}(\theta,\varphi) [Y_\varphi^{\ell'
m'}(\theta,\varphi)]^* \right] \nonumber \\
& & \qquad \qquad = \ell (\ell +1) \delta
_{\ell \ell'}\delta _{m m'}.
\end{eqnarray}
Here $d\Omega_2= \sin \theta \, d\theta \, d\varphi$ denotes the area element on the unit sphere ${\cal S}^2$. We also recall that the gauge-invariant master functions $\psi_{\ell m} (t,r)$ appearing in Eq.~(\ref{ChampE_even}) can be written in the form
\begin{equation}\label{TF_psi}
\psi_{\ell m} (t,r) = \frac{1}{\sqrt{2\pi}} \int_{-\infty}^{+\infty} d\omega \, \psi_{\omega \ell m} (r) e^{-i\omega t}
\end{equation}
where their Fourier components $\psi_{\omega \ell m} (r) $ satisfy the Regge-Wheeler equation
\begin{equation} \label{ZM EQ_Fourier}
\left[\frac{d^2}{d
r_\ast^2} + \omega^2 - V_\ell (r)  \right] \psi_{\omega \ell m} (r) = S_{\omega \ell m} (r).
\end{equation}
Here, $S_{\omega \ell m} (r)$ is a source term, $r_\ast$ denotes the tortoise coordinate which is defined in terms of the radial Schwarzschild coordinate $r$ by $dr/dr_\ast=f(r)$ and is given by $r_\ast(r)=r+2M \ln[r/(2M)-1]$ while
\begin{equation} \label{pot R-W}
V_\ell(r)=f(r)\left(\frac{\ell(\ell+1)}{r^2}\right)
\end{equation}
denotes the Regge-Wheeler potential.

As far as the source term $S_{\omega \ell m} (r)$ appearing in the right-hand side (r.h.s.)~of the Regge-Wheeler equation (\ref{ZM EQ_Fourier}) is concerned, it depends on the components, in the basis of vector spherical harmonics, of the current associated with the charged particle \cite{Folacci:2018vtf}. Its expression can be derived from Eqs.~(\ref{geodesic_equations}) and (\ref{tildeEand gamma}) and we obtain
\begin{equation} \label{SourceRad_gen}
S_{\omega \ell m} (r) =[Y^{\ell m}(\pi/2,0)]^\ast \widetilde{S}_{\omega} (r) e^{+i \omega t_p(r) }
\end{equation}
where
\begin{widetext}
\begin{subequations}  \label{SourceRad_omR}
\begin{eqnarray} \label{SourceRad_omR_a}
& & \widetilde{S}_{\omega} (r) =\frac{q}{\sqrt{2 \pi}}f(r) \left[ + i \omega \frac{r}{(\gamma^{2}-1)r + 2M} - \frac{M \gamma}{\sqrt{r} \, \left[(\gamma^{2}-1)r+2M\right]^{3/2}} \right]
\end{eqnarray}
and with
\begin{eqnarray}
\label{trajectory_Rad}
& & \frac{t_{p}(r)}{2M}=-\frac{2}{3} \left(\frac{r}{2M} \right)^{3/2} -2 \left(\frac{r}{2M} \right)^{1/2} + \ln \left(\frac{\sqrt{\frac{r}{2M}} +1}{\sqrt{\frac{r}{2M}} -1} \right) +  \frac{t_{0}}{2M}
\end{eqnarray}
for the particle starting at rest from infinity (i.e., for $\gamma=1$) and
\begin{eqnarray}
\label{trajectory_RadRel}
& & \frac{t_{p}(r)}{2M}=-\frac{\gamma}{(\gamma^{2}-1)^{3/2}}\sqrt{\left[(\gamma^{2}-1)\frac{r}{2M}\right]
\left[(\gamma^{2}-1)\frac{r}{2M}+1\right]} \nonumber \\
& & \qquad \qquad - \frac{\gamma (2\gamma^{2}-3)}{(\gamma^{2}-1)^{3/2}} \ln\left[\sqrt{(\gamma^{2}-1)\frac{r}{2M}}  + \sqrt{(\gamma^{2}-1)\frac{r}{2M}+1}\right]  \nonumber \\
& & \qquad \qquad+ \ln\left[\frac{\gamma \sqrt{\frac{r}{2M}}+ \sqrt{(\gamma^{2}-1)\frac{r}{2M}+1}}{ \gamma \sqrt{\frac{r}{2M}}- \sqrt{(\gamma^{2}-1)\frac{r}{2M}+1}}\right] +  \frac{t_{0}}{2M}
\end{eqnarray}
\end{subequations}
\end{widetext}
for a particle projected with a finite kinetic energy at infinity (i.e., for $\gamma > 1$). In Eqs.~(\ref{trajectory_Rad}) and (\ref{trajectory_RadRel}), $t_0$ is an arbitrary integration constant.

\subsection{Regge-Wheeler equation and ${\cal S}$-matrix}
\label{SecIIb}

The Regge-Wheeler equation (\ref{ZM EQ_Fourier}) can be solved by using the machinery of Green's functions (see, e.g., Ref.~\cite{Breuer:1974uc} for its use in the context of BH physics). {\it Mutatis mutandis}, taking into account Eq.~(\ref{SourceRad_gen}), the reasoning of Sec.~IIC of Ref.~\cite{Folacci:2018cic} permits us to obtain the asymptotic expression, for $r \to +\infty$, of the partial amplitudes $\psi_{\omega\ell m}(r)$. We have
\begin{subequations}\label{Partial_Response}
\begin{equation}
\label{Partial_Response_a}
\psi_{\omega\ell m}(r)= e^{+i \omega r_\ast(r) } \, \frac{K[\ell,\omega]}{2i\omega A^{(-)}_\ell (\omega)} \,
[Y^{\ell m}(\pi/2,0)]^\ast
\end{equation}
with
\begin{eqnarray}
\label{Partial_Response_b}
& & K[\ell,\omega] =  \int_{2M}^{+\infty} \frac{dr'}{f(r')} \,\phi_{\omega, \ell}^\mathrm {in}(r')
\, \widetilde{S}_{\omega} (r') e^{i \omega t_p(r') }.
\end{eqnarray}
\end{subequations}
Here, we have introduced the solution $\phi_{\omega, \ell}^\mathrm {in} (r) $ of the homogeneous Regge-Wheeler equation
\begin{equation}
\label{H_RW_equation}
\left[\frac{d^{2}}{dr_{\ast}^{2}}+\omega^{2}-V_{\ell}(r)\right] \phi_{\omega, \ell}^\mathrm {in}= 0
\end{equation}
which is defined by its behavior at the event horizon $r=2M$ (i.e., for $r_\ast \to -\infty$) and at spatial infinity $r \to +\infty$ (i.e., for $r_\ast \to +\infty$):
\begin{eqnarray}
\label{bc_in}
& & \phi_{\omega, \ell}^{\mathrm {in}}(r_{*}) \sim \left\{
\begin{aligned}
&\!\!\displaystyle{e^{-i\omega r_\ast}}  \,\, (r_\ast \to -\infty)\\
&\!\!\displaystyle{A^{(-)}_\ell (\omega) e^{-i\omega r_\ast} + A^{(+)}_\ell (\omega) e^{+i\omega r_\ast}} \,\, (r_\ast \to +\infty).
\end{aligned}
\right. \nonumber\\
&&
\end{eqnarray}
The coefficients $A^{(-)}_\ell (\omega)$ and  $A^{(+)}_\ell (\omega)$ appearing in Eqs.~(\ref{Partial_Response}) and (\ref{bc_in}) are complex amplitudes. By evaluating, first for $r_\ast \to - \infty$ and then for $r_\ast \to + \infty$, the Wronskian involving the function $\phi_{\omega\ell}^{\mathrm {in}}$ and its complex conjugate, we can show that they are linked by
\begin{equation}\label{Rel_conserv_Apm}
|A^{(-)}_\ell (\omega)|^2 - |A^{(+)}_\ell (\omega)|^2 = 1.
\end{equation}
Moreover, with the numerical calculation of the Maxwell scalar $\phi_2$ as well as the study of its properties in mind, it is important to note that
\begin{subequations}\label{Sym_om}
\begin{eqnarray}
& & \phi_{-\omega, \ell}^{\mathrm {in}} (r)=  \left[\phi_{\omega, \ell}^{\mathrm {in}}(r)  \right]^\ast, \label{Sym_om_a}\\
& & A^{(\pm )}_\ell (-\omega) =  [A^{( \pm)}_\ell (\omega) ]^\ast. \label{Sym_om_b}
\end{eqnarray}
\end{subequations}

It is worth pointing out that the boundary conditions (\ref{bc_in}) for $\phi_{\omega, \ell}^\mathrm {in} (r) $ and therefore the expression  (\ref{Partial_Response}) of the partial amplitudes $\psi_{\omega\ell m} (r)$ involve the ${\cal S}$-matrix defined by (see, e.g., Ref.~\cite{DeWitt:2003pm})
\begin{eqnarray}
\label{S_matrix_def}
& & {\cal S}_\ell (\omega) = \left(\,
\begin{aligned}
&\!\!\displaystyle{\qquad\quad 1/A^{(-)}_\ell (\omega)} &  \displaystyle{A^{(+)}_\ell (\omega)/A^{(-)}_\ell (\omega)} \\
&\!\!-\displaystyle{[A^{(+)}_\ell (\omega)]^\ast/A^{(-)}_\ell (\omega)} & \displaystyle{1/A^{(-)}_\ell (\omega)}\quad
\end{aligned}
\right). \nonumber\\
&&
\end{eqnarray}
Due to Eq.~(\ref{Sym_om_b}), this matrix satisfies the symmetry property ${\cal S}_\ell (-\omega)=\left[ {\cal S}_\ell (\omega) \right]^\ast$ and, due to Eq.~(\ref{Rel_conserv_Apm}), it is in addition unitary, i.e., it satisfies $ {\cal S} {\cal S}^\dag= {\cal S}^\dag {\cal S}= \mathrm{1}$. Here, it is interesting to recall that, in Eq.~(\ref{S_matrix_def}), the term $1/A^{(-)}_\ell (\omega)$ and the term $A^{(+)}_\ell (\omega)/A^{(-)}_\ell (\omega)$ are, respectively, the transmission coefficient $T_\ell(\omega)$ and the reflection coefficient $R^\mathrm{in}_\ell(\omega)$ corresponding to the scattering problem defined by Eq.~(\ref{bc_in}). As far as the coefficient $-[A^{(+)}_\ell (\omega)]^\ast/A^{(-)}_\ell (\omega)$ is concerned, it can be considered as the reflection coefficient $R^\mathrm{up}_\ell(\omega)$ involved in the scattering problem defining the modes $\phi^\mathrm{up}_{\omega, \ell} (r)$ \cite{DeWitt:2003pm}.

\subsection{Compact expression for the multipole expansion of the Maxwell scalar $\phi_2$}
\label{SecIIc}

We first insert Eq.~(\ref{Partial_Response_a}) into Eq.~(\ref{TF_psi}) and we have
 \begin{subequations}\label{TF_psi_bis}
\begin{equation}
\label{TF_psi_bis_a}
\psi_{\ell m} (t,r) = \psi_\ell (t,r)  \, [Y^{\ell m}(\pi/2,0)]^\ast
\end{equation}
where
\begin{eqnarray}
\label{TF_psi_bis_b}
& & \psi_\ell (t,r) = \frac{1}{\sqrt{2\pi}} \int_{-\infty}^{+\infty} d\omega \, e^{-i\omega[ t-r_\ast(r)]}\, \frac{K[\ell,\omega]}{2i\omega A^{(-)}_\ell (\omega)}. \nonumber \\
& &
\end{eqnarray}
\end{subequations}
We now substitute Eq.~(\ref{TF_psi_bis}) into Eq.~(\ref{ChampE_even}). Furthermore, without loss of generality, we assume that the electromagnetic radiation is observed in a direction lying in the BH equatorial plane and making an angle $\varphi \in [0, \pi]$ with the trajectory of the particle (due to symmetry considerations, we can restrict our study to this interval). By then using the addition theorem for scalar spherical harmonics in the form
\begin{equation}\label{ThAd_HS}
\sum_{m=-\ell}^{+\ell}    Y^{\ell m}(\theta,\varphi) [Y^{\ell m}(\pi/2,0)]^\ast = \frac{2\ell +1}{4 \pi} P_\ell (\sin\theta \cos\varphi)
\end{equation}
where $P_\ell (x)$ denotes the Legendre polynomial of degree $\ell$ \cite{AS65}, we obtain, for $r\to +\infty$,
\begin{equation}\label{hc_def}
r\, E_\theta  (t,r,\theta=\pi/2,\varphi)=0
\end{equation}
and
\begin{eqnarray}\label{hp_def}
& & r\, E_\varphi  (t,r,\theta=\pi/2,\varphi)= -\frac{1}{\sqrt{2\pi}} \int_{-\infty}^{+\infty} d\omega \, e^{-i\omega[ t-r_\ast(r)]} \nonumber \\
& & \qquad \times \left[\frac{1}{4\pi}\sum_{\ell=1}^{+\infty}\frac{2\ell+1}{\ell(\ell+1)}\, \frac{K[\ell,\omega]}{2 A^{(-)}_\ell (\omega)}  \, W_\ell (\cos\varphi) \right].
\end{eqnarray}
In Eq.~(\ref{hp_def}), we have introduced the angular function
\begin{equation}\label{ang_function}
W_\ell (\cos\varphi) = \frac{\partial}{\partial \varphi} P_\ell(\cos\varphi).
\end{equation}
Finally, taking into account Eq.~(\ref{Relations_ChampB_even_odd}), we can write by inserting Eqs.~(\ref{hc_def}) and (\ref{hp_def}) into Eq.~(\ref{phi2_def})
\begin{eqnarray}\label{phi2_ExpressionDef}
& & \frac{\sqrt{2} \, r}{i}\, \phi_2 (t,r,\theta=\pi/2,\varphi)= \frac{1}{\sqrt{2 \pi}} \int_{-\infty}^{+\infty} d\omega \, e^{-i\omega[ t-r_\ast(r)]} \nonumber \\
& & \qquad\qquad \times \left[ \frac{1}{4\pi} \sum_{\ell=1}^{+\infty} \frac{2\ell+1}{\ell(\ell+1)} \, \frac{K[\ell,\omega]}{2 A^{(-)}_\ell (\omega)}  \, W_\ell (\cos\varphi)\right]
\end{eqnarray}
for $r \to + \infty$.

\subsection{Regularization of the partial waveform amplitudes $\psi_{\omega\ell m}$ and the Maxwell scalar $\phi_2$ }
\label{SecIId}

To construct the Maxwell scalar $\phi_2$, we need to regularize the partial amplitudes $\psi_{\omega\ell m}(r)$ or, more precisely, $K[\ell, \omega]$. Indeed, the partial waveforms~(\ref{Partial_Response}) as integrals over the radial Schwarzschild coordinate are divergent at infinity. This is due to the behavior of the source~(\ref{SourceRad_omR}) for $r \to \infty$.

To regularize $K[\ell, \omega]$, we integrate twice by parts and use the homogeneous Regge-Wheeler equation~\eqref{H_RW_equation}. Then, by dropping intentionally the boundary terms  at $r \to \infty$ (regularization), we obtain
\begin{subequations}
\begin{eqnarray}
\label{Partial_Response_c}
& & K[\ell,\omega] = q \, \ell(\ell+1) \frac{\widetilde{K}[\ell,\omega]}{i \omega}
\end{eqnarray}
with
\begin{eqnarray}
\label{Partial_Response_d}
& & \widetilde{K}[\ell,\omega] = \frac{1}{\sqrt{2\pi}} \int_{2M}^{+\infty} dr' \,\phi_{\omega, \ell}^\mathrm {in}(r')\, \frac{ e^{i \omega t_p(r') }}{r'^{2}}.
\end{eqnarray}
\end{subequations}

Now, by inserting Eqs.~\eqref{Partial_Response_c} and \eqref{Partial_Response_d} into Eq.~\eqref{phi2_ExpressionDef}, we obtain for the Maxwell scalar
\begin{eqnarray}\label{phi2_ExpressionDef_reg}
& & \frac{\sqrt{2} \, r}{i q}\, \phi_2 (t,r,\theta=\pi/2,\varphi)= \frac{1}{\sqrt{2\pi}} \int_{-\infty}^{+\infty} d\omega \, e^{-i\omega[ t-r_\ast(r)]} \nonumber \\
& & \qquad\qquad \times \left[ \sum_{\ell=1}^{+\infty} \frac{2\ell+1}{4\pi} \, \frac{\widetilde{K}[\ell,\omega]}{2 i \omega A^{(-)}_\ell (\omega)}  \,W_\ell (\cos\varphi) \right].
\end{eqnarray}
It is in addition interesting to note that, by inserting Eqs.~\eqref{Partial_Response_c} and \eqref{Partial_Response_d} into the partial waveform amplitudes~(\ref{Partial_Response_a}), we can recover the amplitude term derived by Cardoso, Lemos and Yoshida in Ref.~\cite{Cardoso:2003cn} working in the Zerilli gauge \cite{Zerilli:1971wd,Zerilli:1974ai}.

Moreover, with the numerical calculation of the Maxwell scalar $\phi_2$ as well as the study of its properties in mind, we can observe that
\begin{subequations}\label{Sym_om_cd}
\begin{equation}
\widetilde{K}[\ell,-\omega] =  \left[\widetilde{K}[\ell,\omega] \right]^\ast \label{Sym_om_c}
\end{equation}
and
\begin{equation}
\widetilde{K}[\ell,-\omega]/A^{(-)}_\ell (-\omega) =  \left[ \widetilde{K}[\ell,\omega]/A^{(-)}_\ell (\omega) \right]^\ast \label{Sym_om_d}
\end{equation}
\end{subequations}
as a consequence of Eqs.~(\ref{Sym_om_a}) and (\ref{Sym_om_b}).  Due to relation (\ref{Sym_om_d}), we can see that the term in square brackets in Eq.~(\ref{phi2_ExpressionDef_reg}) satisfies the Hermitian symmetry property and, as a consequence, that the Maxwell scalar $\phi_2$ is a purely imaginary quantity. Similarly, it is worth pointing out that the electromagnetic field is a real quantity.

\subsection{Two alternative expressions for the multipole expansion of the Maxwell scalar $\phi_2$}
\label{SecIIe}

It is important to realize that Eq.~(\ref{phi2_ExpressionDef_reg}) can also be written as
\begin{eqnarray}\label{phi2_ExpressionDef_reg_P}
& & \frac{\sqrt{2} \, r}{i q}\, \phi_2 (t,r,\theta=\pi/2,\varphi)= \frac{1}{\sqrt{2\pi}} \int_{-\infty}^{+\infty} d\omega \, e^{-i\omega[ t-r_\ast(r)]} \nonumber \\
& & \qquad\qquad \times \left[ \sum_{\ell=0}^{+\infty} \frac{2\ell+1}{4\pi} \, \frac{\widetilde{K}[\ell,\omega]}{2 i \omega A^{(-)}_\ell (\omega)}  \,W_\ell (\cos\varphi) \right].
\end{eqnarray}
Indeed, it is possible to start at $\ell=0$ the discrete sum over $\ell$ by noting that
\begin{equation}\label{Mode_Zero}
W_0 (\cos\varphi)=\frac{\partial}{\partial\varphi} P_0(\cos \varphi)= 0
\end{equation}
and that we have formally
\begin{equation}\label{Formally_ell0}
 A_0^{(-)}(\omega)=1 \quad \mathrm{and} \quad \widetilde{K}[0,\omega] \,\, \mathrm{regular}.
\end{equation}
These last two results are due to the fact that, for $\ell=0$, the solution of the problem (\ref{H_RW_equation})--(\ref{bc_in}) is $\phi_{\omega, 0}^\mathrm {in} (r) = e^{-i\omega r_\ast}$ because the Regge-Wheeler potential (\ref{pot R-W}) vanishes. Of course, in general, it is more natural to work with the multipole expansion (\ref{phi2_ExpressionDef_reg}) of the Maxwell scalar $\phi_2$ but, in Sec.~\ref{SecIIIc}, we shall take (\ref{phi2_ExpressionDef_reg_P}) as a departure point because it will permit us to use the Poisson summation formula in its standard form.

Similarly, it is important to note that Eq.~(\ref{phi2_ExpressionDef_reg_P}) can be rewritten in the form
\begin{eqnarray}\label{phi2_ExpressionDef_reg_SW}
& & \frac{\sqrt{2} \, r}{i q}\, \phi_2 (t,r,\theta=\pi/2,\varphi)= \frac{1}{\sqrt{2\pi}} \int_{-\infty}^{+\infty} d\omega \, e^{-i\omega[ t-r_\ast(r)]} \nonumber \\
& &\, \times \left[ \sum_{\ell=0}^{+\infty} (-1)^{\ell} \frac{2\ell+1}{4\pi} \, \frac{\widetilde{K}[\ell,\omega]}{2 i \omega A^{(-)}_\ell (\omega)}  \,W_\ell (-\cos\varphi) \right].
\end{eqnarray}
Indeed, we can recover Eq.~(\ref{phi2_ExpressionDef_reg_P}) from Eq.~(\ref{phi2_ExpressionDef_reg_SW}) by using the relation \cite{AS65}
\begin{equation}\label{prop_Pell}
P_\ell (- \cos \varphi) = (-1)^\ell P_\ell (\cos \varphi)
\end{equation}
in connection with the definition (\ref{ang_function}). In Sec.~\ref{SecIIId}, we shall take Eq.~(\ref{phi2_ExpressionDef_reg_SW}) as a departure point because it will permit us to use the Sommerfeld-Watson transform in its standard form.

\subsection{Quasinormal ringdown associated with the Maxwell scalar $\phi_2$}
\label{SecIIf}

  The quasinormal ringdown  $\phi^\text{\tiny{QNM}}_2$ generated by the charged particle falling radially from infinity into a Schwarzschild BH can be
extracted from Eq.~\eqref{phi2_ExpressionDef_reg} by following, \textit{mutatis mutandis}, the reasoning of Sec.~II~E of Ref.~\cite{Folacci:2018sef}. We then obtain
 \begin{eqnarray}
\label{response_QNM}
& & \frac{\sqrt{2} \, r}{iq}\, \phi^\text{\tiny{QNM}}_2 (t,r,\theta=\pi/2,\varphi)=
 - 2 \sqrt{2\pi} \, \operatorname{Re}   \left[ \, \sum^{+\infty}_{   \ell =1} \sum^{+\infty}_{n =1} \phantom{\frac{\widetilde{K}}{A_{\ell}^{(+)}}}  \right. \nonumber \\
& & \,\, \left.\frac{2\ell+1}{4\pi} \, {\cal{B}}_{\ell n} \frac{\widetilde{K}[\ell,\omega_{\ell n}]}{A_{\ell}^{(+)}(\omega_{\ell n})}\, e^{-i \omega_{\ell n}[t-r_\ast(r)]}  \, W_\ell (\cos \varphi)\right]
\end{eqnarray}
where
\begin{equation}
\label{excitation_factor_QNM}
{\cal{B}}_{\ell n}=\left[\frac{1}{2 \omega}\,\,\frac{A_{\ell}^{(+)}(\omega)}{\frac{d}{d \omega}A_{\ell}^{(-)}(\omega)}\right]_{\omega=\omega_{\ell n}}
\end{equation}
denotes the excitation factor associated with the $(\ell,n)$ quasinormal mode (QNM) of complex frequency $\omega_{\ell n}$. Its expression involves the residue of the function $1/A^{(-)}_\ell (\omega)$ at $\omega=\omega_{\ell n}$. It should be noted that Eq.~(\ref{response_QNM}) has been obtained by gathering the contributions of the quasinormal frequencies $\omega_{\ell n}$ and $-\omega_{\ell n}^\ast$ taking into account the relations (\ref{Sym_om_b}) and (\ref{Sym_om_c}) which remain valid in the complex $\omega $ plane. As a consequence, the quasinormal ringdown waveform $\phi^\text{\tiny{QNM}}_2$ appears clearly as a purely imaginary quantity.

Let us finally recall that, due to the exponentially divergent behavior of the terms $e^{-i \omega_{\ell n}[t-r_\ast(r)]}$ as $t$ decreases, the ringdown waveform $\phi^\text{\tiny{QNM}}_2$ does not provide
physically relevant results at early times. It is therefore necessary to determine, from physical considerations, a starting time $t_\mathrm{start}$ for the BH ringdown. In general, by taking $t_\mathrm{start}=t_p(3M)$, i.e., the moment the particle crosses the photon sphere, we can obtain physically relevant results.

\section{Maxwell scalar $\phi_2$, its CAM representations and its Regge pole approximations}
\label{SecIII}

In this section, we shall derive two exact CAM representations of the Maxwell scalar $\phi_2$, the first one by using the Poisson summation formula~\cite{MorseFeshbach1953} and the second one by working with the Sommerfeld-Watson transformation~\cite{Watson18,Sommerfeld49,Newton:1982qc}. These representations can be written as (the Fourier transform of) a sum over Regge poles plus background integrals along the positive real axis and the imaginary axis of the CAM plane. We shall also consider the Regge pole part of these representations as approximations of the Maxwell scalar $\phi_2$ which can be evaluated numerically from the Regge trajectories followed by the Regge poles and by the excitation factors of the associated Regge modes.

In order to construct the two CAM representations of the Maxwell scalar $\phi_2$ and the associated Regge pole approximations, we shall follow, \textit{mutatis mutandis}, Sec.~III of Ref.~\cite{Folacci:2018sef}.

\subsection{Some preliminary remarks concerning analytic extensions in the CAM plane}
\label{SecIIIa}

\begin{figure*}
\centering
 \includegraphics[scale=0.55]{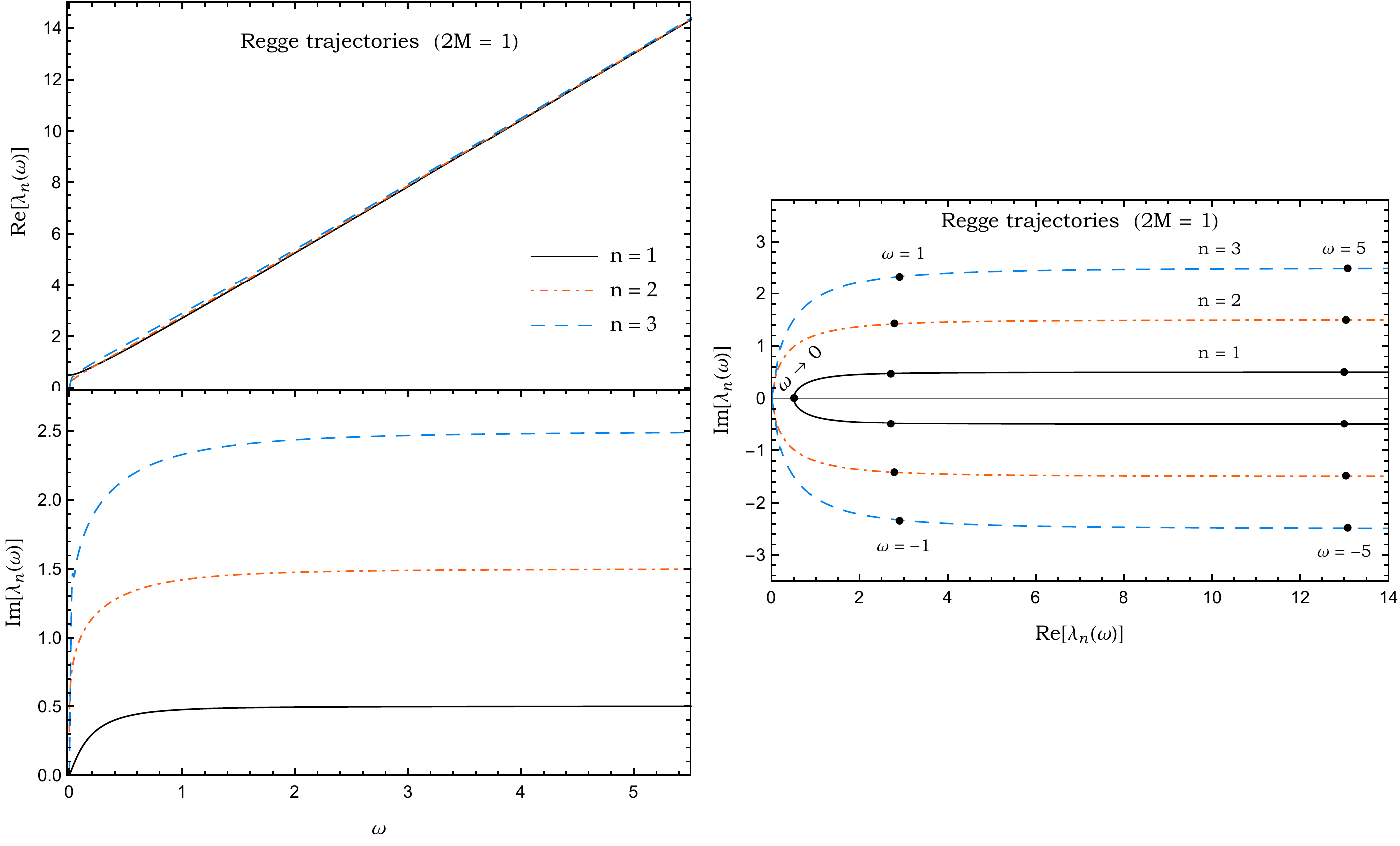}
\caption{\label{TR_PR_S_1} Regge trajectories of the first three Regge poles corresponding to electromagnetism in the Schwarzschild BH ($2M=1$). The relation (\ref{PR_chgt_om}) permits us to describe the Regge trajectories for $\omega<0$ by noting that $\operatorname{Re} [\lambda_n(\omega)]$ and $\operatorname{Im} [\lambda_n(\omega)]$ are, respectively, even and odd functions of $\omega$. We observe, in particular, the migration of the Regge poles in the CAM plane.}
\end{figure*}

\begin{figure}
\centering
 \includegraphics[scale=0.55]{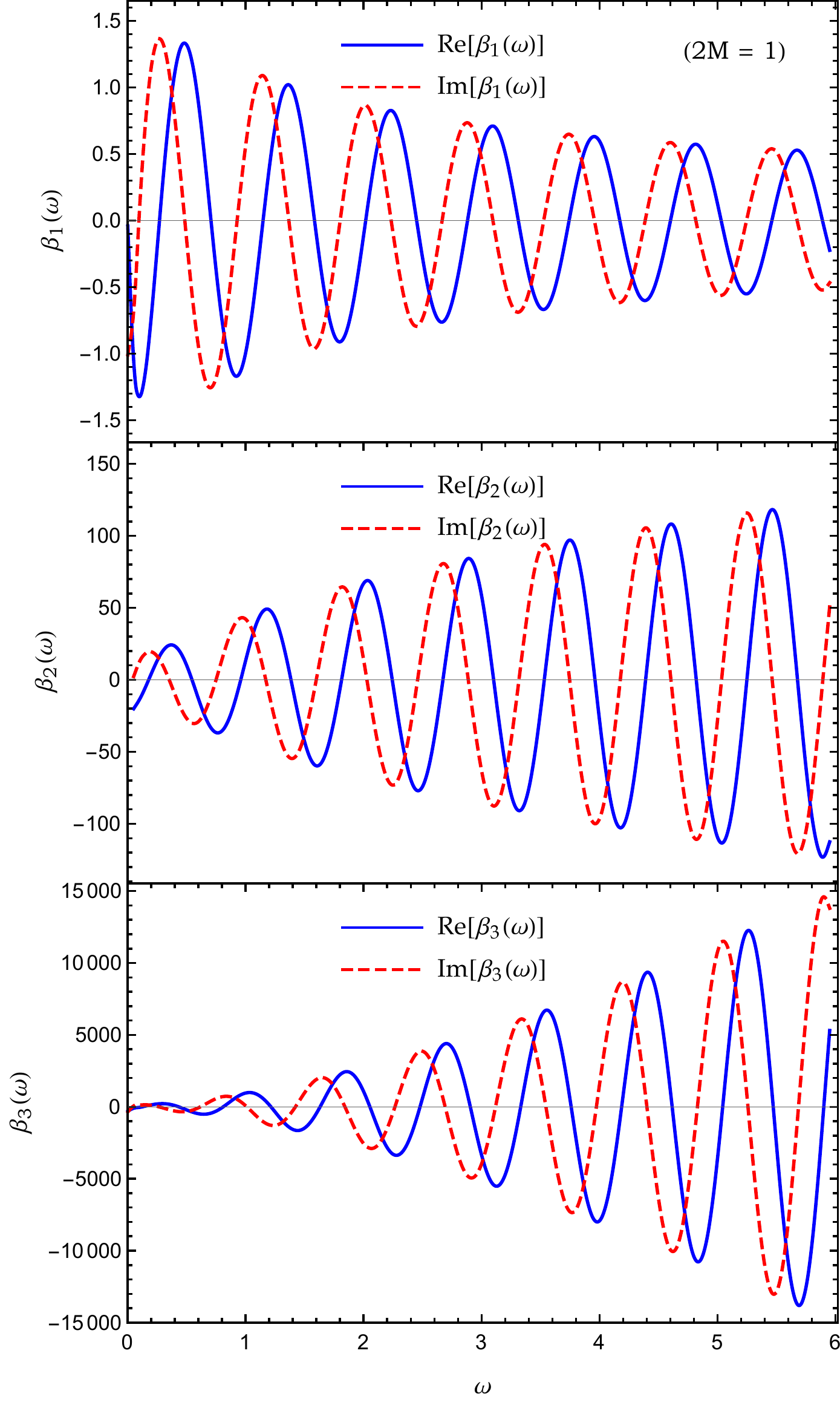}
\caption{\label{RM_excitation_factors_S_1} Regge trajectories of the Regge-mode excitation factors ($2M=1$). We consider the Regge modes corresponding to the first three Regge poles of which the behavior has been displayed in Fig.~\ref{TR_PR_S_1}. The relation (\ref{EF_PR_chgt_om}) permits us to describe the Regge trajectories for $\omega<0$ by noting that $\operatorname{Re} [\beta_n(\omega)]$ and $\operatorname{Im} [\beta_n(\omega)]$ are, respectively, odd and even functions of $\omega$.}
\end{figure}

The CAM machinery permitting us to derive the CAM representations of the multipolar waveform $\phi_2$ requires us to replace in Eqs.~(\ref{phi2_ExpressionDef_reg_P}) and (\ref{phi2_ExpressionDef_reg_SW}) the angular momentum $\ell \in \mathbb{N}$ by the angular momentum $\lambda = \ell +1/2 \in \mathbb{C}$ and therefore to work into the CAM plane. As a consequence, we need to have at our disposal the functions $W_{\lambda -1/2} (\cos \varphi)$, $W_{\lambda -1/2} (-\cos \varphi)$, $A^{(\pm)}_{\lambda-1/2} (\omega)$ and $\widetilde{K}[\lambda-1/2,\omega]$ which are ``the'' analytic extensions of $W_{\ell} (\cos \varphi)$, $W_{\ell} (-\cos \varphi)$, $A^{(\pm)}_{\ell} (\omega)$ and $\widetilde{K}[\ell,\omega]$ in the complex $\lambda$ plane. We recall that the uniqueness problem for such analytic extensions is a difficult problem. We have briefly discussed it in Sec.~IIIA of Ref.~\cite{Folacci:2018sef} (see also Chap.~13 of Ref.~\cite{Newton:1982qc}). Here, in order to construct these analytic extensions, we shall adopt minimal prescriptions that will be justified by the results we shall obtain in Sec.~\ref{SecIV}.

The angular functions $W_{\ell} (\cos \varphi)$, $W_{\ell} (-\cos \varphi)$ are defined from the Legendre polynomial $P_\ell (z)$ [see Eq.~(\ref{ang_function})] of which the analytic extension usually considered is the hypergeometric function \cite{AS65}
\begin{equation}\label{Def_ext_LegendreP}
P_{\lambda -1/2} (z) = F(1/2-\lambda,1/2+\lambda;1;(1-z)/2].
\end{equation}
As a consequence, it is natural to take
\begin{eqnarray}\label{AngFunction_W_ext}
& & W_{\lambda -1/2} (\pm \cos \varphi) \nonumber \\
& & \qquad = \frac{\partial}{\partial \varphi} F(1/2-\lambda,1/2+\lambda;1;(1 \mp \cos \varphi)/2]
\end{eqnarray}
and it is worth noting that, due to the properties of the hypergeometric function, we have
\begin{equation}\label{prop_ext_W_a}
W_{-\lambda -1/2} (\pm \cos \varphi) = W_{\lambda -1/2} (\pm \cos \varphi)
\end{equation}
and
\begin{equation}\label{prop_ext_W_b}
W_{\lambda -1/2} (\pm \cos \varphi) = [W_{\lambda^\ast -1/2} (\pm \cos \varphi)]^\ast.
\end{equation}
Here, it is crucial to keep in mind that, while the angular functions $W_\ell (\pm \cos \phi)$ are well defined for $\varphi \in [0,\pi]$, this is not the case for their analytic extensions $W_{\lambda -1/2} (\pm \cos \phi)$. Indeed, due to the pathologic behavior of $P_{\lambda -1/2} (z)$ at $z=-1$, $W_{\lambda -1/2} (\cos \phi)$ diverges in the limit $\varphi \to \pi$ and $W_{\lambda -1/2} (-\cos \phi)$ diverges in the limit $\varphi \to 0$. Due to the problems they generate on the Regge pole approximations of $\phi_2$, we shall return to these results later.

Analytic extensions $A^{(\pm)}_{\lambda-1/2} (\omega)$ and $\widetilde{K}[\lambda-1/2,\omega]$ of $A^{(\pm)}_{\ell} (\omega)$ and $\widetilde{K}[\ell,\omega]$ are obtained by assuming that the function $\phi_{\omega, \lambda-1/2}^\mathrm {in} (r) $ and the coefficients $A^{(\pm)}_{\lambda-1/2} (\omega)$ can be defined by the problem (\ref{H_RW_equation})--(\ref{bc_in}) where now $\ell \in \mathbb{N}$ is replaced by $\lambda-1/2 \in \mathbb{C}$. Such prescription permits us, in particular, to extend in the CAM plane the properties (\ref{Sym_om_a}), (\ref{Sym_om_b}), (\ref{Sym_om_c}) and (\ref{Sym_om_d}). In the following, we shall therefore consider that
\begin{subequations}\label{Sym_om_CAM_ab}
\begin{eqnarray}
& & \phi_{-\omega, \lambda -1/2}^{\mathrm {in}} (r)=  [\phi_{\omega, \lambda^\ast -1/2}^{\mathrm {in}}(r)]^\ast, \label{Sym_om_CAM_a}\\
& & A^{(\pm )}_{\lambda -1/2} (-\omega) =  [A^{( \pm)}_{\lambda^\ast -1/2} (\omega) ]^\ast, \label{Sym_om_CAM_b}
\end{eqnarray}
\end{subequations}
and that
\begin{subequations}\label{Sym_om_CAM_cd}
\begin{eqnarray}
& & \widetilde{K}[\lambda -1/2,-\omega] =  \left[ \widetilde{K}[\lambda^\ast -1/2,\omega] \right]^\ast, \label{Sym_om_CAM_c} \\
& & \widetilde{K}[\lambda -1/2,-\omega]/A^{(-)}_{\lambda -1/2} (-\omega) =  \nonumber \\
& & \qquad\qquad\qquad \left[ \widetilde{K}[\lambda^\ast -1/2,\omega]/A^{(-)}_{\lambda^\ast -1/2} (\omega) \right]^\ast. \label{Sym_om_CAM_d}
\end{eqnarray}
\end{subequations}

\subsection{Regge poles, Regge modes and associated excitation factors}
\label{SecIIIb}

In the next two subsections, contour deformations in the CAM plane will permit us to collect, by using Cauchy's residue theorem, the contributions from the Regge poles of the ${\cal S}$-matrix or, more precisely, from the poles, in the complex $\lambda$ plane and for $\omega \in \mathbb{R}$, of the matrix ${\cal S}_{\lambda-1/2}(\omega)$. It should be noted that these poles can be defined as the zeros $\lambda_n(\omega)$ with $n=1, 2, 3, \dots  $ and $\omega \in \mathbb{R}$ of the coefficient $A^{(-)}_{\lambda-1/2} (\omega)$ [see Eq.~(\ref{S_matrix_def})]. They therefore satisfy
\begin{equation}\label{PR_def_Am}
A^{(-)}_{\lambda_n(\omega)-1/2} (\omega)=0.
\end{equation}

The Regge poles corresponding to electromagnetism in the Schwarzschild BH have been studied in Refs.~\cite{Decanini:2009mu,Dolan:2009nk}. It should be recalled that, for $\omega >0$, the Regge poles lie in the first and third quadrants of the CAM plane, symmetrically distributed with respect to the origin $O$ of this plane. In this article, due to the use of Fourier transforms, we must be able to locate the Regge poles even for $\omega < 0$. In fact, from the symmetry relation (\ref{Sym_om_CAM_b}), we have
\begin{equation}\label{PR_chgt_om}
\lambda_n(-\omega)=[\lambda_n(\omega)]^\ast
\end{equation}
and we can see immediately that, for $\omega < 0$, the Regge poles lie in the second and fourth quadrants of the CAM plane, symmetrically distributed with respect to the origin $O$ of this plane. Moreover, if we consider the Regge trajectories $\lambda_n(\omega)$ with $\omega \in ]-\infty,+\infty[$, we can observe the migration of the Regge poles. More precisely, as $\omega $ decreases, the Regge poles lying in the first (third) quadrant of the CAM plane migrate in the fourth (second) one.

It should be noted that the solutions of the problem (\ref{H_RW_equation})--(\ref{bc_in}) with $\ell$ replaced by $\lambda_n(\omega)-1/2 $ are modes that are purely outgoing at infinity and purely ingoing at the horizon. They are the ``Regge modes'' of the Schwarzschild BH \cite{Decanini:2009mu,Dolan:2009nk}. Because of the analogy with the QNMs, it is natural to define excitation factors for these modes. In fact, they will appear in the CAM representations of the Maxwell scalar $\phi_2$. By analogy with the excitation factor associated with the $(\ell,n)$ QNM of complex frequency $\omega_{\ell n}$ [see Eq.~(\ref{excitation_factor_QNM})], we define the excitation factor of the Regge mode associated with the Regge pole $\lambda_n(\omega)$ by
\begin{equation}\label{excitation_factor_RP}
\beta_n(\omega)=\left[\frac{1}{2 \omega}\,\,\frac{A_{\lambda -1/2}^{(+)}(\omega)}{\frac{d}{d \lambda}A_{\lambda -1/2}^{(-)}(\omega)}\right]_{\lambda=\lambda_n(\omega)}.
\end{equation}
Its expression involves the residue of the matrix ${\cal S}_{\lambda-1/2}(\omega)$ [or, more precisely, of the function $1/A^{(-)}_{\lambda-1/2} (\omega)$] at $\lambda=\lambda_n(\omega)$. It should be noted that, due to Eq.~(\ref{Sym_om_CAM_b}), we have
\begin{equation}\label{EF_PR_chgt_om}
\beta_n(-\omega)=-[\beta_n(\omega)]^\ast.
\end{equation}

We have displayed the Regge trajectories of the first three Regge poles as well as the Regge trajectories of the corresponding excitation factors in Figs.~\ref{TR_PR_S_1} and \ref{RM_excitation_factors_S_1}. These numerical results have been obtained by using, {\it mutatis mutandis}, the methods that have permitted us to obtain, in Refs.~\cite{Folacci:2018vtf,Folacci:2018cic}, for the electromagnetic field and for gravitational waves, the complex quasinormal frequencies of the QNMs and the associated excitation factors (see, e.g., Sec.~IV~A of Ref.~\cite{Folacci:2018cic}).

It is important to point out that, in Refs.~\cite{Decanini:2002ha,Decanini:2009mu}, we have established a connection between the Regge modes and the (weakly damped) QNMs of the Schwarzschild BH. It will play a central role in the interpretation of our results in Sec.~\ref{SecIV}, and we recall that, for a given $n$, the Regge trajectory $\lambda_n(\omega)$ with $\omega \in \mathbb{R}$ encodes information on the complex quasinormal frequencies $\omega_{\ell n}$ with $\ell=1, 2, 3, \dots$ In fact, the index $n=1, 2, 3, \dots  $ not only permits us to distinguish between the different Regge poles but is also associated with the family of quasinormal frequencies generated by the Regge modes.

\subsection{CAM representation and Regge pole approximation of the Maxwell scalar $\phi_2$ based on the Poisson summation formula}
\label{SecIIIc}

The first CAM representation of the Maxwell scalar $\phi_2$ can be derived from Eq.~(\ref{phi2_ExpressionDef_reg_P}) by using the Poisson summation formula~\cite{MorseFeshbach1953} as well as Cauchy's residue theorem. This can be achieved by following, \textit{mutatis mutandis}, the reasoning of Sec.~III~C of Ref.~\cite{Folacci:2018sef} which has permitted us to construct a CAM representation of the Weyl scalar $\Psi_4$. In fact, it is even possible to avoid repeating in detail this reasoning: indeed, we can note that Eq.~(24) of Ref.~\cite{Folacci:2018sef} defining $\Psi_4$ and which is the departure of the reasoning of Sec.~IIIC of Ref.~\cite{Folacci:2018sef} and Eq.~(\ref{phi2_ExpressionDef_reg_P}) of the present article are related by the correspondences
\begin{subequations}\label{PSI4-phi2}
\begin{eqnarray}
r \, \Psi_4 (t,r,\theta=\pi/2,\varphi) & \longleftrightarrow & \frac{\sqrt{2} \, r}{iq} \, \phi_2 (t,r,\theta=\pi/2,\varphi), \nonumber \\
& & \label{PSI4-phi2_a}\\
\frac{i \omega K[\ell,\omega]}{4 A^{(-)}_\ell (\omega)} & \longleftrightarrow &  \frac{\widetilde{K}[\ell,\omega]}{2 i \omega A^{(-)}_\ell (\omega)}, \label{PSI4-phi2_b} \\
Z_\ell (\cos \varphi) & \longleftrightarrow &  W_\ell (\cos \varphi). \label{PSI4-phi2_c}
\end{eqnarray}
\end{subequations}
As a consequence, Eqs.~(48) and (49) of Ref.~\cite{Folacci:2018sef} which provide a CAM representation of the Weyl scalar $\Psi_4$ can be translated to obtain directly a CAM representation of the Maxwell scalar $\phi_2$. We can write
\begin{widetext}
\begin{equation}\label{CAM_phi2_ExpressionDef_Ptot}
\phi_2 (t,r,\theta=\pi/2,\varphi)= \phi^{\text{\tiny{B}} \, \textit{\tiny{(P)}}}_2 (t,r,\theta=\pi/2,\varphi) + \phi^{\text{\tiny{RP}} \, \textit{\tiny{(P)}}}_2 (t,r,\theta=\pi/2,\varphi)
\end{equation}
where
\begin{subequations}\label{CAM_phi2_ExpressionDef_P}
\begin{eqnarray}\label{CAM_phi2_ExpressionDef_P_Background}
& & \frac{\sqrt{2} \, r}{iq}\, \phi^{\text{\tiny{B}} \, \textit{\tiny{(P)}}}_2 (t,r,\theta=\pi/2,\varphi)= \frac{1}{\sqrt{2\pi}} \int_{-\infty}^{+\infty} d\omega \, e^{-i\omega[ t-r_\ast(r)]} \left[ \int_{0}^{\infty} d\lambda \, \frac{\lambda}{2\pi} \,  \frac{\widetilde{K}[\lambda-1/2,\omega]}{2i\omega A^{(-)}_{\lambda-1/2} (\omega)}  \, W_{\lambda -1/2} (\cos \varphi) \right. \nonumber \\
& & \qquad \qquad\qquad \left. -\frac{1}{4\pi} \int_{0}^{+i\infty} d\lambda \, \frac{\lambda e^{i\pi \lambda}}{\cos (\pi \lambda)}  \frac{\widetilde{K}[\lambda-1/2,\omega]}{2i\omega A^{(-)}_{\lambda-1/2} (\omega)}  \, W_{\lambda -1/2} (\cos \varphi)   \right.  \nonumber \\
& & \qquad \qquad\qquad \left.  -\frac{1}{4\pi} \int_{0}^{-i\infty} d\lambda \, \frac{\lambda e^{-i \pi \lambda}}{\cos (\pi \lambda)}  \frac{\widetilde{K}[\lambda-1/2,\omega]}{2i\omega A^{(-)}_{\lambda-1/2} (\omega)}  \, W_{\lambda -1/2} (\cos \varphi)   \right]
\end{eqnarray}
is a background integral contribution and where
\begin{eqnarray}\label{CAM_phi2_ExpressionDef_P_RP}
& & \frac{\sqrt{2} \, r}{iq}\, \phi^{\text{\tiny{RP}} \, \textit{\tiny{(P)}}}_2 (t,r,\theta=\pi/2,\varphi)=\frac{1}{\sqrt{2\pi}} \int_{-\infty}^{+\infty} d\omega \, e^{-i\omega[ t-r_\ast(r)]} \nonumber \\
& & \qquad \times \left[ - {\cal H}(\omega)  \sum_{n=1}^{+\infty}  \frac{\lambda_n(\omega) \beta_n(\omega) e^{i\pi \lambda_n(\omega)}}{\cos [\pi \lambda_n(\omega)]} \, \frac{\widetilde{K}[\lambda_n(\omega)-1/2,\omega]}{2 A^{(+)}_{\lambda_n(\omega)-1/2} (\omega)} \,  W_{\lambda_n(\omega) -1/2} (\cos \varphi)   \right. \nonumber \\
& & \qquad \quad \left. + {\cal H}(-\omega) \sum_{n=1}^{+\infty}  \frac{\lambda_n(\omega) \beta_n(\omega) e^{-i\pi \lambda_n(\omega)}}{\cos [\pi \lambda_n(\omega)]} \, \, \frac{\widetilde{K}[\lambda_n(\omega)-1/2,\omega]}{2 A^{(+)}_{\lambda_n(\omega)-1/2} (\omega)} \,  W_{\lambda_n(\omega) -1/2} (\cos \varphi)    \right]
\end{eqnarray}
\end{subequations}
\end{widetext}
is the Fourier transform of a sum over Regge poles. In these expressions, ${\cal H}$ denotes the Heaviside step function and we have introduced the analytic extensions discussed in Sec.~\ref{SecIIIa} as well as the Regge poles and the associated excitation factors considered in Sec.~\ref{SecIIIb}.

We can again check that $\phi_2$ is a purely imaginary quantity by now considering this new expression. Indeed, due to the relations (\ref{prop_ext_W_b}) and (\ref{Sym_om_CAM_d}), the first term as well as the sum of the second and third terms within the square brackets on the r.h.s.~of Eq.~(\ref{CAM_phi2_ExpressionDef_P_Background}) satisfy the Hermitian symmetry property. Such a property is also satisfied by the sum of the two terms within the square bracket on the r.h.s.~of Eq.~(\ref{CAM_phi2_ExpressionDef_P_RP}) as a consequence of the relations (\ref{prop_ext_W_b}), (\ref{Sym_om_CAM_b}), (\ref{Sym_om_CAM_c}), (\ref{PR_chgt_om}) and (\ref{EF_PR_chgt_om}).

Of course, Eqs.~\eqref{CAM_phi2_ExpressionDef_Ptot} and (\ref{CAM_phi2_ExpressionDef_P}) provide an exact representation for the Maxwell scalar $\phi_2$, equivalent to the initial expression \eqref{phi2_ExpressionDef}. From this CAM representation of $\phi_2$, we can extract the contribution denoted by $\phi^{\text{\tiny{RP}} \, \textit{\tiny{(P)}}}_2$ and given by Eq.~(\ref{CAM_phi2_ExpressionDef_P_RP}) which, as a sum over Regge poles, is only an approximation of $\phi_2$. In Sec.~\ref{SecIV}, we shall compare it with the exact expression (\ref{phi2_ExpressionDef}) of $\phi_2$. However, when considering the term $\phi^{\text{\tiny{RP}} \, \textit{\tiny{(P)}}}_2$ alone, we shall encounter some problems due to the pathological behavior of $W_{\lambda_n(\omega) -1/2} (\cos \varphi)$ for $\varphi \to \pi$ (see Sec.~\ref{SecIIIa}). In fact, both the Regge pole approximation $\phi^{\text{\tiny{RP}} \, \textit{\tiny{(P)}}}_2$ and the background integral contribution $\phi^{\text{\tiny{B}} \, \textit{\tiny{(P)}}}_2$ are divergent in the limit $\varphi \to \pi$ but it is worth pointing out that their sum (\ref{CAM_phi2_ExpressionDef_Ptot}) does not present any pathology.

\subsection{CAM representation and Regge pole approximation of the Maxwell scalar $\phi_2$ based on the Sommerfeld-Watson transform}
\label{SecIIId}

The second CAM representation of the Maxwell scalar $\phi_2$ can be derived from Eq.~(\ref{phi2_ExpressionDef_reg_SW}) by using the Sommerfeld-Watson transformation \cite{Watson18,Sommerfeld49,Newton:1982qc} as well as Cauchy's residue theorem. This can be achieved by following, \textit{mutatis mutandis}, the reasoning of Sec.~III~D of Ref.~\cite{Folacci:2018sef} which has permitted us to construct a CAM representation of the Weyl scalar $\Psi_4$. Here again, we avoid repeating in detail this reasoning: we note that Eq.~(26) of Ref.~\cite{Folacci:2018sef} defining $\Psi_4$ and which is the departure of the reasoning of Sec.~III~D of Ref.~\cite{Folacci:2018sef} and Eq.~(\ref{phi2_ExpressionDef_reg_SW}) of the present article are related by the correspondences (\ref{PSI4-phi2_a}), (\ref{PSI4-phi2_b}) and
\begin{equation}\label{PSI4-phi2_d}
Z_\ell (- \cos \varphi)  \longleftrightarrow   W_\ell (- \cos \varphi).
\end{equation}
As a consequence, Eqs.~(52) and (53) of Ref.~\cite{Folacci:2018sef} which provide a CAM representation of the Weyl scalar $\Psi_4$ permit us to obtain directly a CAM representation of the Maxwell scalar $\phi_2$. We have

\begin{widetext}
\begin{equation}\label{CAM_phi2_ExpressionDef_SWtot}
\phi_2 (t,r,\theta=\pi/2,\varphi)= \phi^{\text{\tiny{B}} \, \textit{\tiny{(SW)}}}_2 (t,r,\theta=\pi/2,\varphi) + \phi^{\text{\tiny{RP}} \, \textit{\tiny{(SW)}}}_2 (t,r,\theta=\pi/2,\varphi)
\end{equation}
where
\begin{subequations}\label{CAM_phi2_ExpressionDef_SW}
\begin{eqnarray}\label{CAM_phi2_ExpressionDef_SW_Background}
& &\frac{\sqrt{2} \, r}{iq}\, \phi^{\text{\tiny{B}} \, \textit{\tiny{(SW)}}}_2 (t,r,\theta=\pi/2,\varphi)= \frac{1}{\sqrt{2\pi}} \int_{-\infty}^{+\infty} d\omega \, e^{-i\omega[ t-r_\ast(r)]}
\nonumber\\
& & \qquad\qquad\qquad\qquad \times \left[-\frac{1}{8 \pi} \int_{-i\infty}^{+i\infty} d\lambda \, \frac{\lambda}{\cos (\pi \lambda)} \, \frac{\widetilde{K}[\lambda-1/2,\omega]}{\omega A^{(-)}_{\lambda-1/2} (\omega)} \,  W_{\lambda -1/2} (-\cos \varphi) \right]
\end{eqnarray}
is a background integral contribution and where
\begin{eqnarray}\label{CAM_phi2_ExpressionDef_SW_RP}
& &\frac{\sqrt{2} \, r}{iq}\, \phi^{\text{\tiny{RP}} \, \textit{\tiny{(SW)}}}_2 (t,r,\theta=\pi/2,\varphi)= \frac{1}{\sqrt{2\pi}} \int_{-\infty}^{+\infty} d\omega \, e^{-i\omega[ t-r_\ast(r)]}   \nonumber \\
& & \qquad\qquad\qquad\qquad   \times \left[ \sum_{n=1}^{+\infty}  \frac{\lambda_n(\omega) \beta_n(\omega)}{2i \,\cos[\pi \lambda_n(\omega)]}  \, \frac{ \widetilde{K}[\lambda_n(\omega)-1/2,\omega]}{ A^{(+)}_{\lambda_n(\omega)-1/2} (\omega)} \,  W_{\lambda_n(\omega) -1/2} (-\cos \varphi) \right]
\end{eqnarray}
\end{subequations}
\end{widetext}
is the Fourier transform of a sum over Regge poles.

We can again check that $\phi_2$ is a purely imaginary quantity by now considering this last expression. Indeed, due to the relations (\ref{prop_ext_W_b}) and (\ref{Sym_om_CAM_d}), the term within the square brackets on the r.h.s.~of Eq.~(\ref{CAM_phi2_ExpressionDef_SW_Background}) satisfies the Hermitian symmetry property. Such a property is also satisfied by the term within the square brackets on the r.h.s.~of Eq.~(\ref{CAM_phi2_ExpressionDef_SW_RP}) as a consequence of the relations (\ref{prop_ext_W_b}), (\ref{Sym_om_CAM_b}), (\ref{Sym_om_CAM_c}), (\ref{PR_chgt_om}) and (\ref{EF_PR_chgt_om}).

It is important to note that  Eq.~(\ref{CAM_phi2_ExpressionDef_SWtot}) provides an exact expression for the Maxwell scalar $\phi_2$, equivalent to the initial expression (\ref{phi2_ExpressionDef}) and to the expression (\ref{CAM_phi2_ExpressionDef_P}) obtained from the Poisson summation formula.  From this CAM representation of $\phi_2$, we can extract the contribution denoted by $\phi^{\text{\tiny{RP}} \, \textit{\tiny{(SW)}}}_2$ and given by Eq.~(\ref{CAM_phi2_ExpressionDef_SW_RP}) which, as a sum over Regge poles, is only an approximation of $\phi_2$. In Sec.~\ref{SecIV}, we shall compare it with the exact expression (\ref{phi2_ExpressionDef}) of $\phi_2$ and with the Regge pole approximation $\phi^{\text{\tiny{RP}} \, \textit{\tiny{(P)}}}_2$ obtained in Sec.~\ref{SecIIIc}. However, when considering the term $\phi^{\text{\tiny{RP}} \, \textit{\tiny{(SW)}}}_2$ alone, we shall encounter some problems due to the pathological behavior of  $W_{\lambda_n(\omega) -1/2} (-\cos \varphi)$ for $\varphi \to 0$ (see Sec.~\ref{SecIIIa}). In fact, both the Regge pole approximation $\phi^{\text{\tiny{RP}} \, \textit{\tiny{(SW)}}}_2$ and the background integral contribution $\phi^{\text{\tiny{B}} \, \textit{\tiny{(SW)}}}_2$ are divergent in the limit $\varphi \to 0$ but it is worth pointing out that their sum (\ref{CAM_phi2_ExpressionDef_SWtot}) does not present any pathology.

\section{Comparison of the Maxwell scalar $\phi_2$ with its Regge pole approximations}
\label{SecIV}

In this section, we shall construct numerically the multipolar waveform $\phi_2$ given by Eq.~(\ref{phi2_ExpressionDef}) by summing over a large number of partial modes. This is particularly necessary for the radially infalling relativistic or ultra-relativistic particle. We shall also construct the associated quasinormal ringdown $\phi^\text{\tiny{QNM}}_2$ given by Eq.~(\ref{response_QNM}). We shall then compare these two waveforms with the Regge pole approximations $\phi^{\text{\tiny{RP}} \, \textit{\tiny{(P)}}}_2$ and $\phi^{\text{\tiny{RP}} \, \textit{\tiny{(SW)}}}_2$ respectively given by Eqs.~(\ref{CAM_phi2_ExpressionDef_P_RP}) and (\ref{CAM_phi2_ExpressionDef_SW_RP}) and constructed by considering one or two Regge poles. This will allow us to highlight the benefits of working with the Regge pole approximations of $\phi_2$.

\subsection{Computational methods}
\label{SecIVa}

To construct numerically the Maxwell scalar $\phi_2$ as well as its quasinormal and Regge pole approximations, we use, \textit{mutatis mutandis}, the computational methods developed in Refs.~\cite{Folacci:2018vtf,Folacci:2018cic} which allowed us to describe the electromagnetic field and the gravitational waves generated by a particle plunging from the innermost stable circular orbit into a Schwarzschild BH (see, e.g., Sec.~IV~A of Ref.~\cite{Folacci:2018vtf}).

\begin{figure*}
\centering
 \includegraphics[scale=0.50]{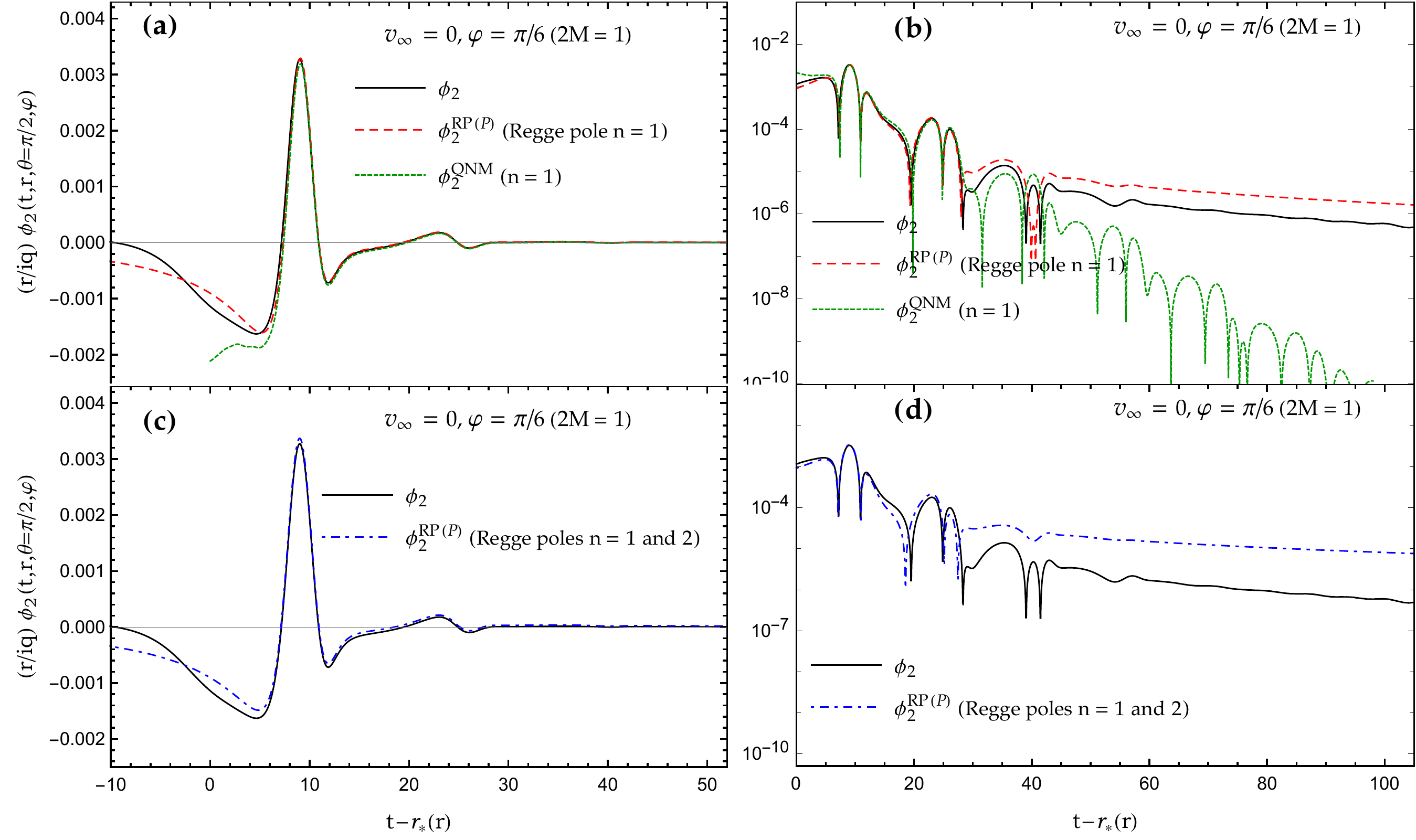}
\caption{\label{P_Exact_QNM_CAM_pis6_v0} The Maxwell scalar $\phi_2$ and its Regge pole approximation $\phi^{\text{\tiny{RP}} \, \textit{\tiny{(P)}}}_2$ for $v_\infty =0$ ($\gamma=1$) and $\varphi=\pi/6$. (a) The Regge pole approximation constructed from only one Regge pole is in very good agreement with the Maxwell scalar $\phi_2$ constructed by summing over the first thirteen partial waves. The associated quasinormal response $\phi^\text{\tiny{QNM}}_2$  obtained by summing over the $(\ell, n)$ QNMs with $n=1$ and $\ell=1, \dots, 13$ is also displayed. At intermediate timescales, it matches very well the Regge pole approximation. (b) Semilog graph corresponding to (a) and showing that the Regge pole approximation describes very well the ringdown, correctly the pre-ringdown phase and roughly the waveform tail. (c) and (d) Taking into account an additional Regge pole does not improve the Regge pole approximation.}
\end{figure*}

\begin{figure*}
\centering
 \includegraphics[scale=0.50]{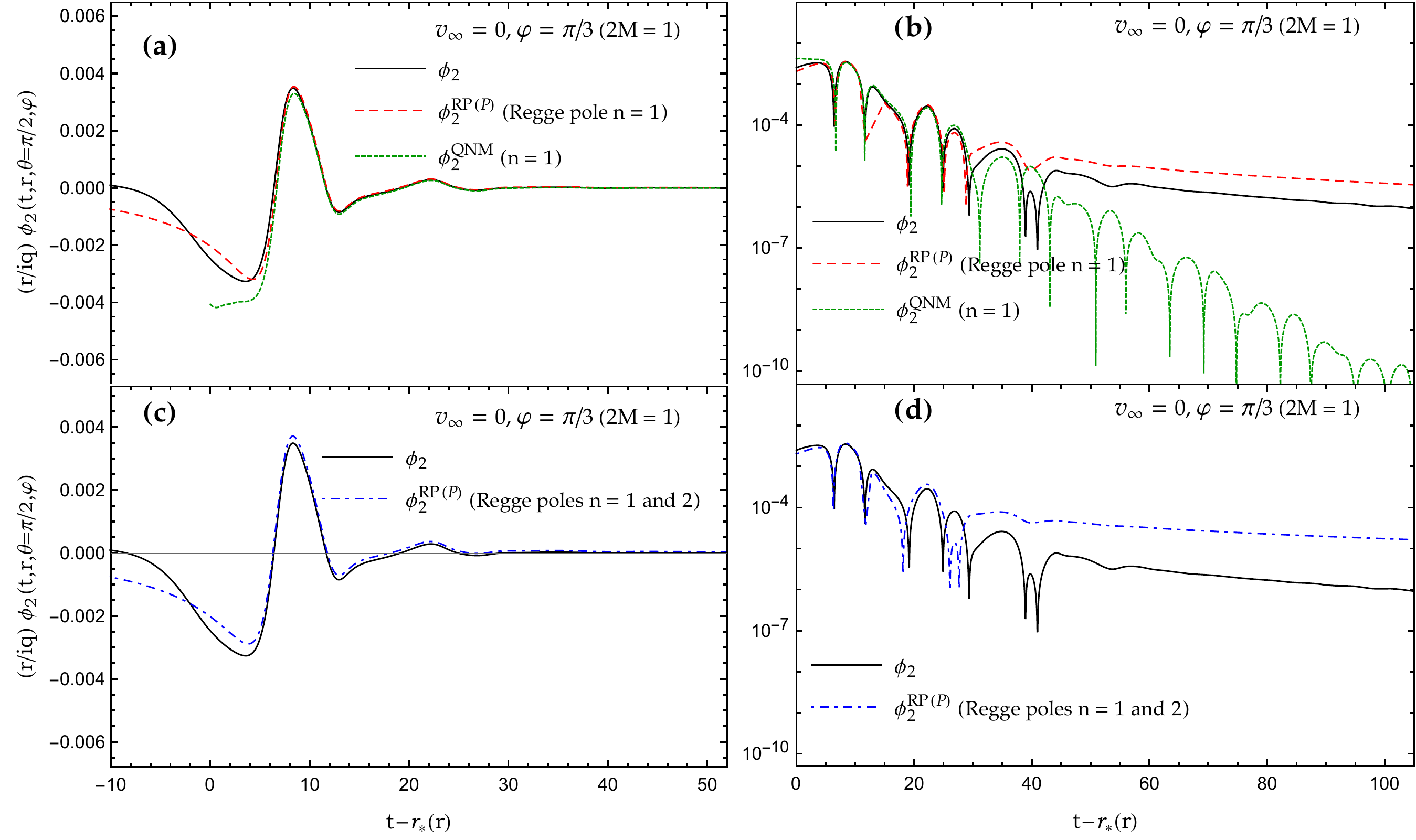}
\caption{\label{P_Exact_QNM_CAM_pis3_v0}  The Maxwell scalar $\phi_2$ and its Regge pole approximation $\phi^{\text{\tiny{RP}} \, \textit{\tiny{(P)}}}_2$ for $v_\infty =0$ ($\gamma=1$) and $\varphi=\pi/3$. (a) The Regge pole approximation constructed from only one Regge pole is in very good agreement with the Maxwell scalar $\phi_2$ constructed by summing over the first thirteen partial waves. The associated quasinormal response $\phi^\text{\tiny{QNM}}_2$  obtained by summing over the $(\ell, n)$ QNMs with $n=1$ and $\ell=1, \dots, 13$ is also displayed. At intermediate timescales, it matches very well the Regge pole approximation. (b) Semilog graph corresponding to (a) and showing that the Regge pole approximation describes very well the ringdown, correctly the pre-ringdown phase and roughly the waveform tail. (c) and (d) Taking into account an additional Regge pole does not improve the Regge pole approximation.}
\end{figure*}

\begin{figure*}
\centering
 \includegraphics[scale=0.50]{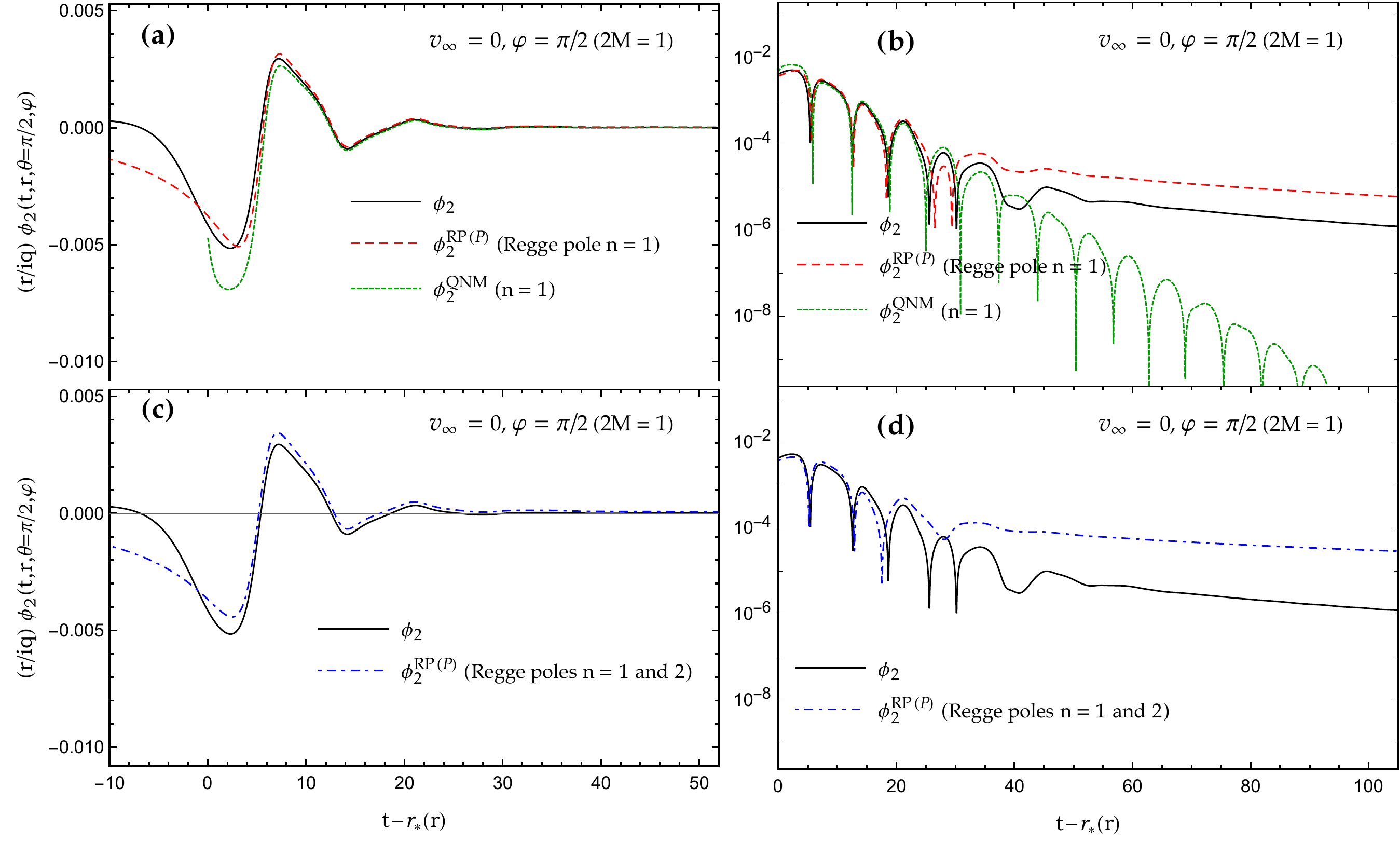}
\caption{\label{P_Exact_QNM_CAM_pis2_v0} The Maxwell scalar $\phi_2$ and its Regge pole approximation $\phi^{\text{\tiny{RP}} \, \textit{\tiny{(P)}}}_2$ for $v_\infty =0$ ($\gamma=1$) and $\varphi=\pi/2$. (a) The Regge pole approximation constructed from only one Regge pole is in rather good agreement with the Maxwell scalar $\phi_2$ constructed by summing over the first thirteen partial waves. The associated quasinormal response $\phi^\text{\tiny{QNM}}_2$  obtained by summing over the $(\ell, n)$ QNMs with $n=1$ and $\ell=1, \dots, 13$ is also displayed. At intermediate timescales, it matches very well the Regge pole approximation. (b) Semilog graph corresponding to (a) and showing that the Regge pole approximation describes very well a large part of the ringdown, correctly the pre-ringdown phase and roughly the waveform tail. (c) and (d) Taking into account an additional Regge pole does not improve the Regge pole approximation.}
\end{figure*}

\begin{figure*}
\centering
 \includegraphics[scale=0.50]{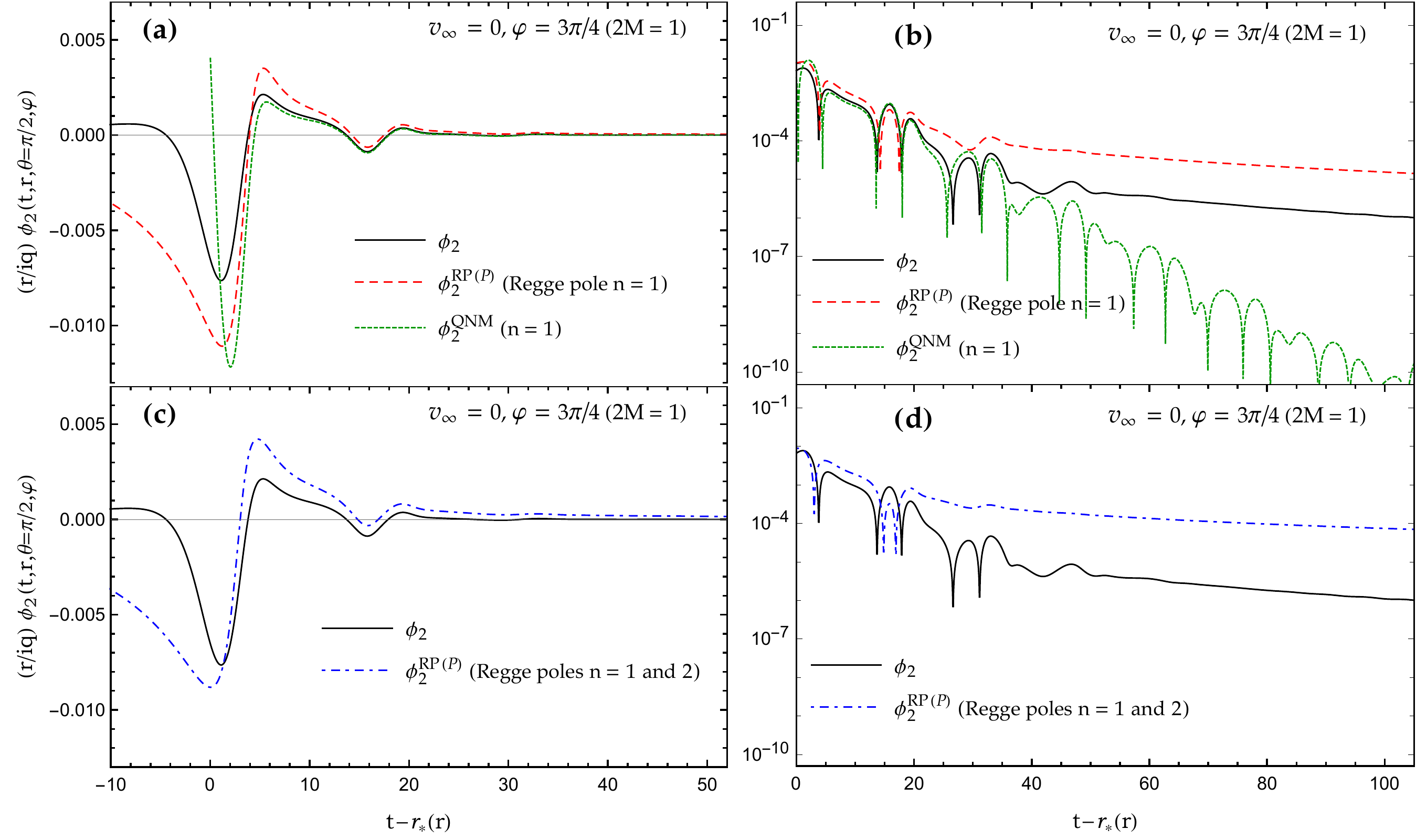}
\caption{\label{P_Exact_QNM_CAM_3pis4_v0} The Maxwell scalar $\phi_2$ and its Regge pole approximation $\phi^{\text{\tiny{RP}} \, \textit{\tiny{(P)}}}_2$ for $v_\infty =0$ ($\gamma = 1$) and $\varphi=3\pi/4$. (a) The Regge pole approximation constructed from only one Regge pole does not match correctly the Maxwell scalar $\phi_2$ constructed by summing over the first thirteen partial waves. The associated quasinormal response $\phi^\text{\tiny{QNM}}_2$  obtained by summing over the $(\ell, n)$ QNMs with $n=1$ and $\ell=1, \dots, 13$ is also displayed. The discrepancy with the Regge pole approximation is obvious. (b) Semilog graph corresponding to (a) and showing that the Regge pole approximation describes correctly a small part of the ringdown. (c) and (d) Taking into account an additional Regge pole does not improve the Regge pole approximation.}
\end{figure*}

 \begin{figure*}
\centering
 \includegraphics[scale=0.49]{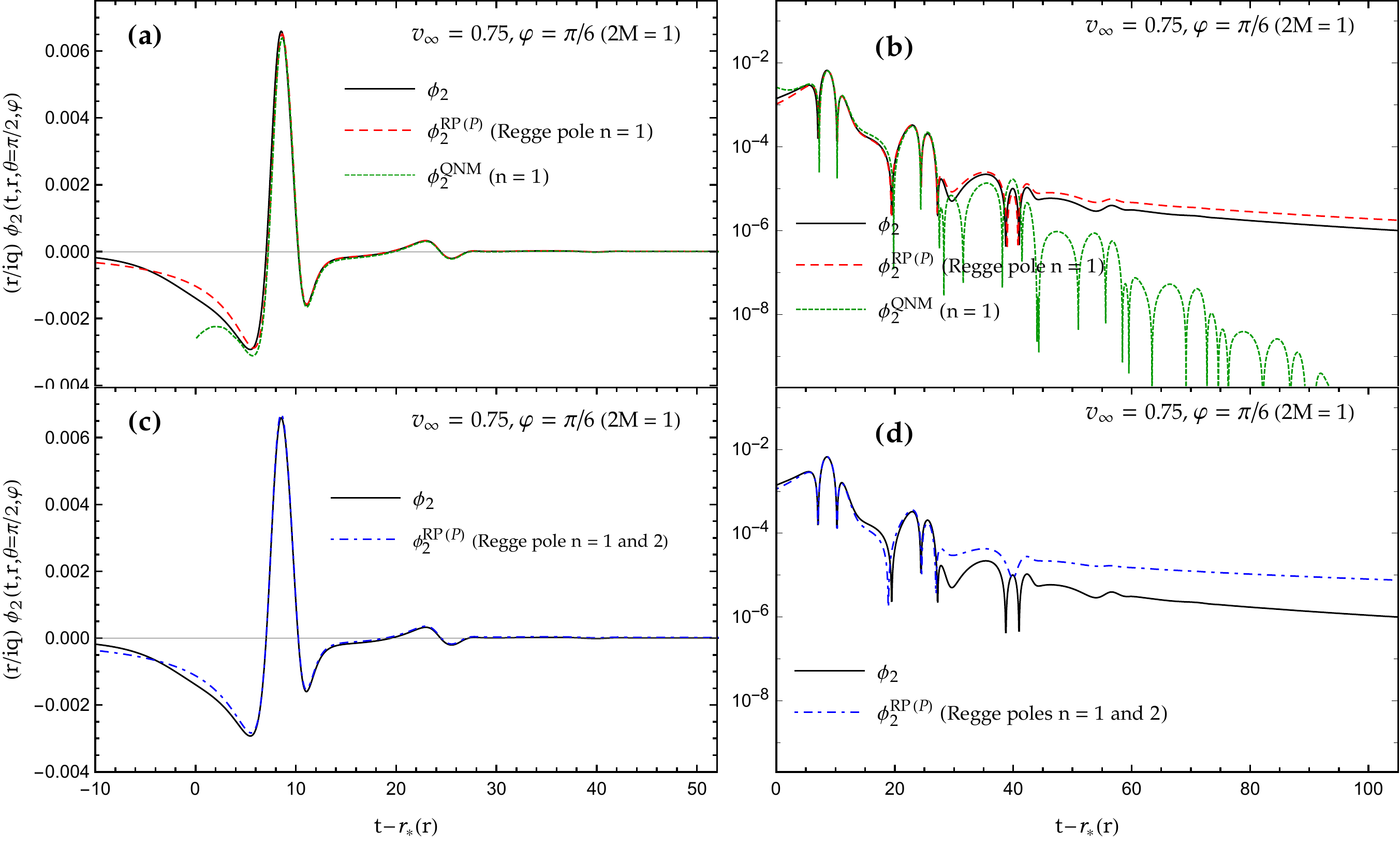}
\caption{\label{P_Exact_QNM_CAM_pis6_v075} The Maxwell scalar $\phi_2$ and its Regge pole approximation $\phi^{\text{\tiny{RP}} \, \textit{\tiny{(P)}}}_2$ for $v_\infty =0.75$ ($\gamma \approx 1.51$) and $\varphi=\pi/6$. (a) The Regge pole approximation constructed from only one Regge pole is in very good agreement with the Maxwell scalar $\phi_2$ constructed by summing over the first fifteen partial waves. The associated quasinormal response $\phi^\text{\tiny{QNM}}_2$  obtained by summing over the $(\ell, n)$ QNMs with $n=1$ and $\ell=1, \dots, 15$ is also displayed. At intermediate timescales, it matches very well the Regge pole approximation. (b) Semilog graph corresponding to (a) and showing that the Regge pole approximation describes very well the pre-ringdown and ringdown phases and correctly approximates the waveform tail. (c) and (d) Taking into account an additional Regge pole does not improve the Regge pole approximation.}
\end{figure*}

 \begin{figure*}
\centering
 \includegraphics[scale=0.49]{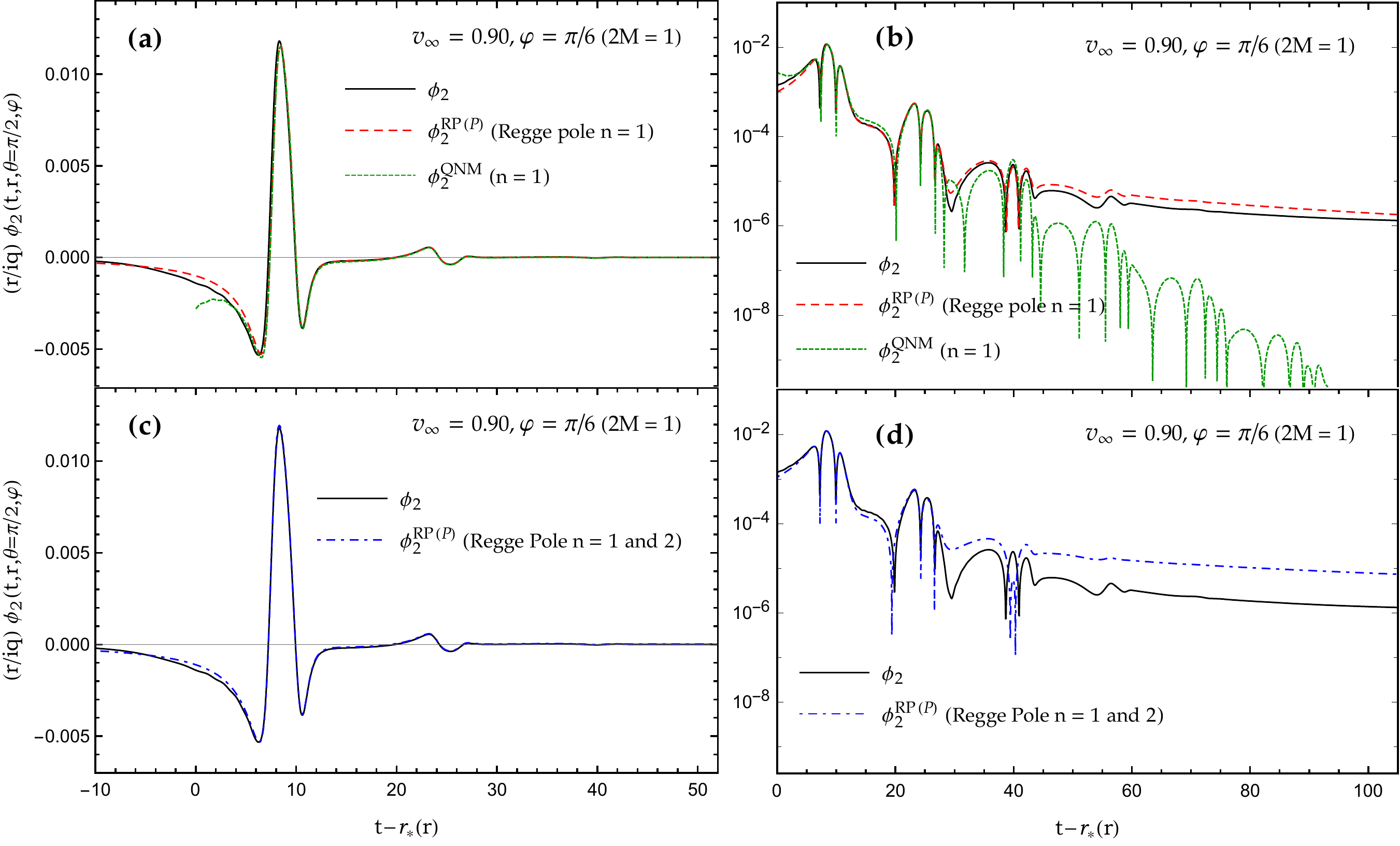}
\caption{\label{P_Exact_QNM_CAM_pis6_v090} The Maxwell scalar $\phi_2$ and its Regge pole approximation $\phi^{\text{\tiny{RP}} \, \textit{\tiny{(P)}}}_2$ for $v_\infty =0.90$ ($\gamma \approx 2.29$) and $\varphi=\pi/6$. (a) The Regge pole approximation constructed from only one Regge pole is in very good agreement with the Maxwell scalar $\phi_2$ constructed by summing over the first fifteen partial waves. The associated quasinormal response $\phi^\text{\tiny{QNM}}_2$  obtained by summing over the $(\ell, n)$ QNMs with $n=1$ and $\ell=1, \dots, 15$ is also displayed. At intermediate timescales, it matches very well the Regge pole approximation. (b) Semilog graph corresponding to (a) and showing that the Regge pole approximation describes very well the whole signal. (c) and (d) Taking into account an additional Regge pole does not improve the Regge pole approximation.}
\end{figure*}

 \begin{figure*}
\centering
 \includegraphics[scale=0.49]{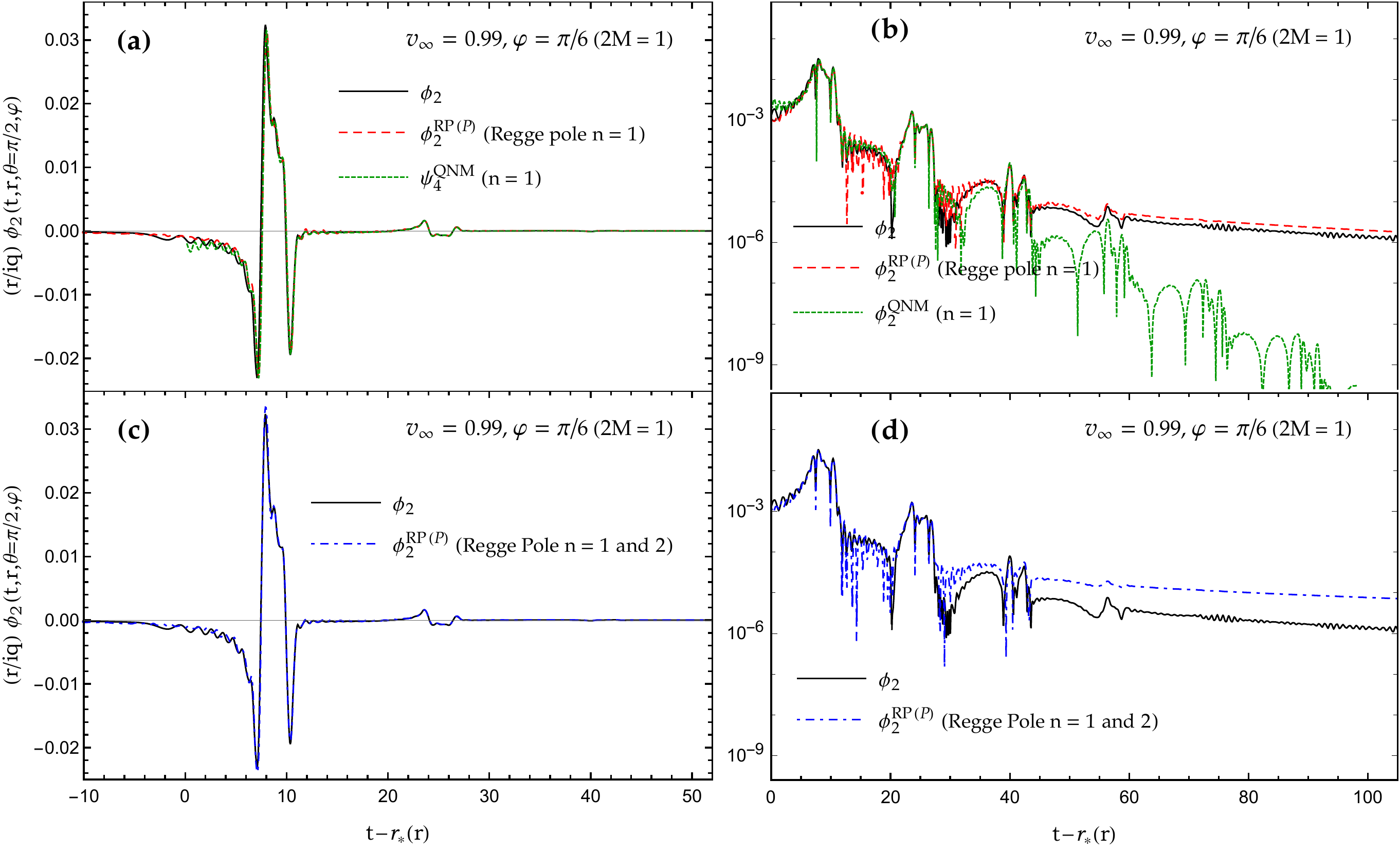}
\caption{\label{P_Exact_QNM_CAM_pis6_v099} The Maxwell scalar $\phi_2$ and its Regge pole approximation $\phi^{\text{\tiny{RP}} \, \textit{\tiny{(P)}}}_2$ for $v_\infty =0.99$ ($\gamma \approx 7.09$) and $\varphi=\pi/6$. (a) The Regge pole approximation constructed from only one Regge pole is in impressive agreement with the Maxwell scalar $\phi_2$ constructed by summing over the first nineteen partial waves. The associated quasinormal response $\phi^\text{\tiny{QNM}}_2$  obtained by summing over the $(\ell, n)$ QNMs with $n=1$ and $\ell=1, \dots, 19$ is also displayed. At intermediate timescales, it matches very well the Regge pole approximation. (b) Semilog graph corresponding to (a) and showing that the Regge pole approximation describes very well the whole signal. (c) and (d) Taking into account an additional Regge pole does not improve the Regge pole approximation.}
\end{figure*}

 \begin{figure*}
\centering
 \includegraphics[scale=0.49]{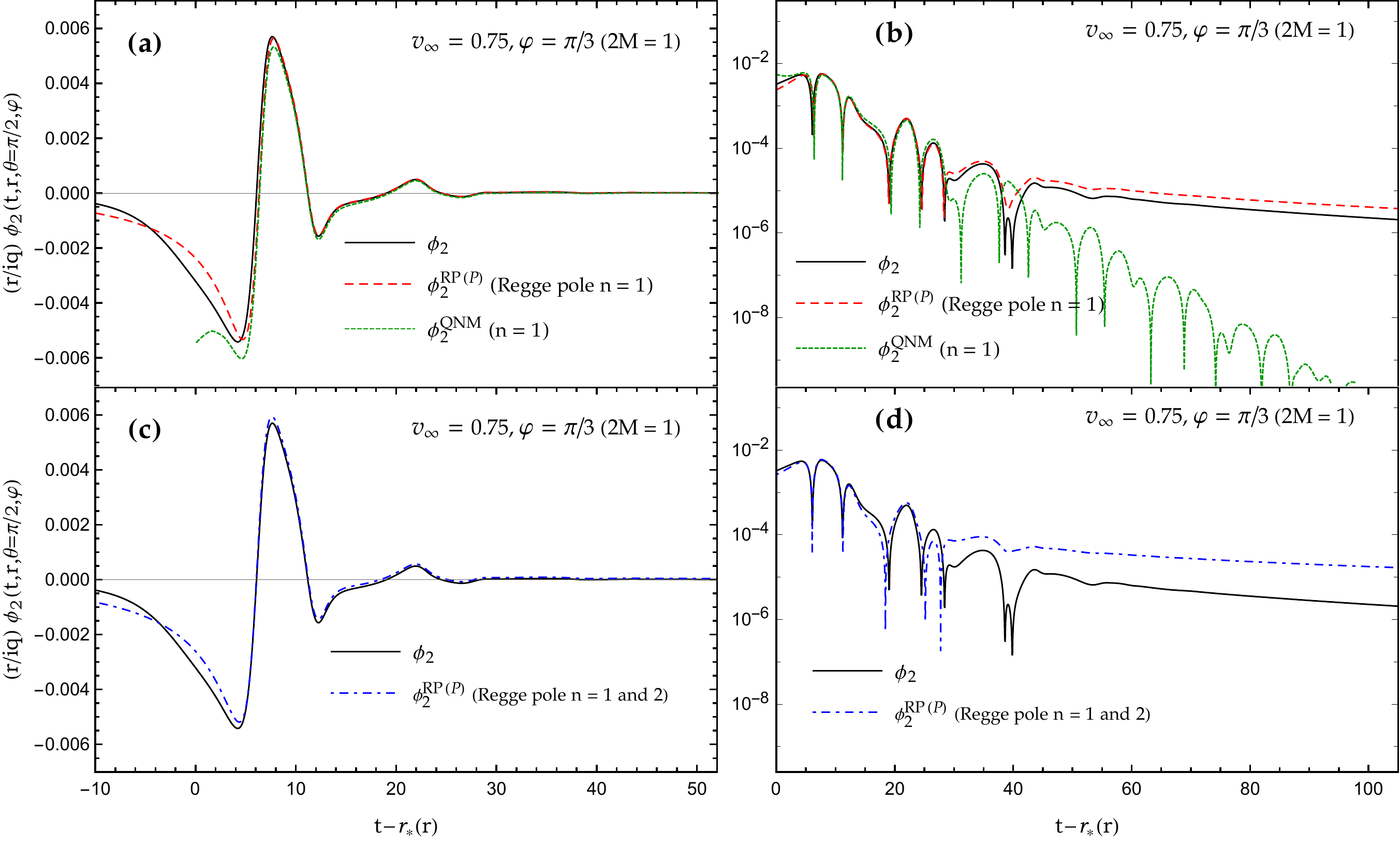}
\caption{\label{P_Exact_QNM_CAM_pis3_v075} The Maxwell scalar $\phi_2$ and its Regge pole approximation $\phi^{\text{\tiny{RP}} \, \textit{\tiny{(P)}}}_2$ for $v_\infty =0.75$ ($\gamma \approx 1.51$) and $\varphi=\pi/3$. (a) The Regge pole approximation constructed from only one Regge pole is in very good agreement with the Maxwell scalar $\phi_2$ constructed by summing over the first fifteen partial waves. The associated quasinormal response $\phi^\text{\tiny{QNM}}_2$  obtained by summing over the $(\ell, n)$ QNMs with $n=1$ and $\ell=1, \dots, 15$ is also displayed. At intermediate timescales, it matches very well the Regge pole approximation. (b) Semilog graph corresponding to (a) and showing that the Regge pole approximation describes very well the ringdown and correctly the pre-ringdown phase and the waveform tail.  (c) and (d) Taking into account an additional Regge pole does not improve the Regge pole approximation.}
\end{figure*}

\begin{figure*}
\centering
 \includegraphics[scale=0.49]{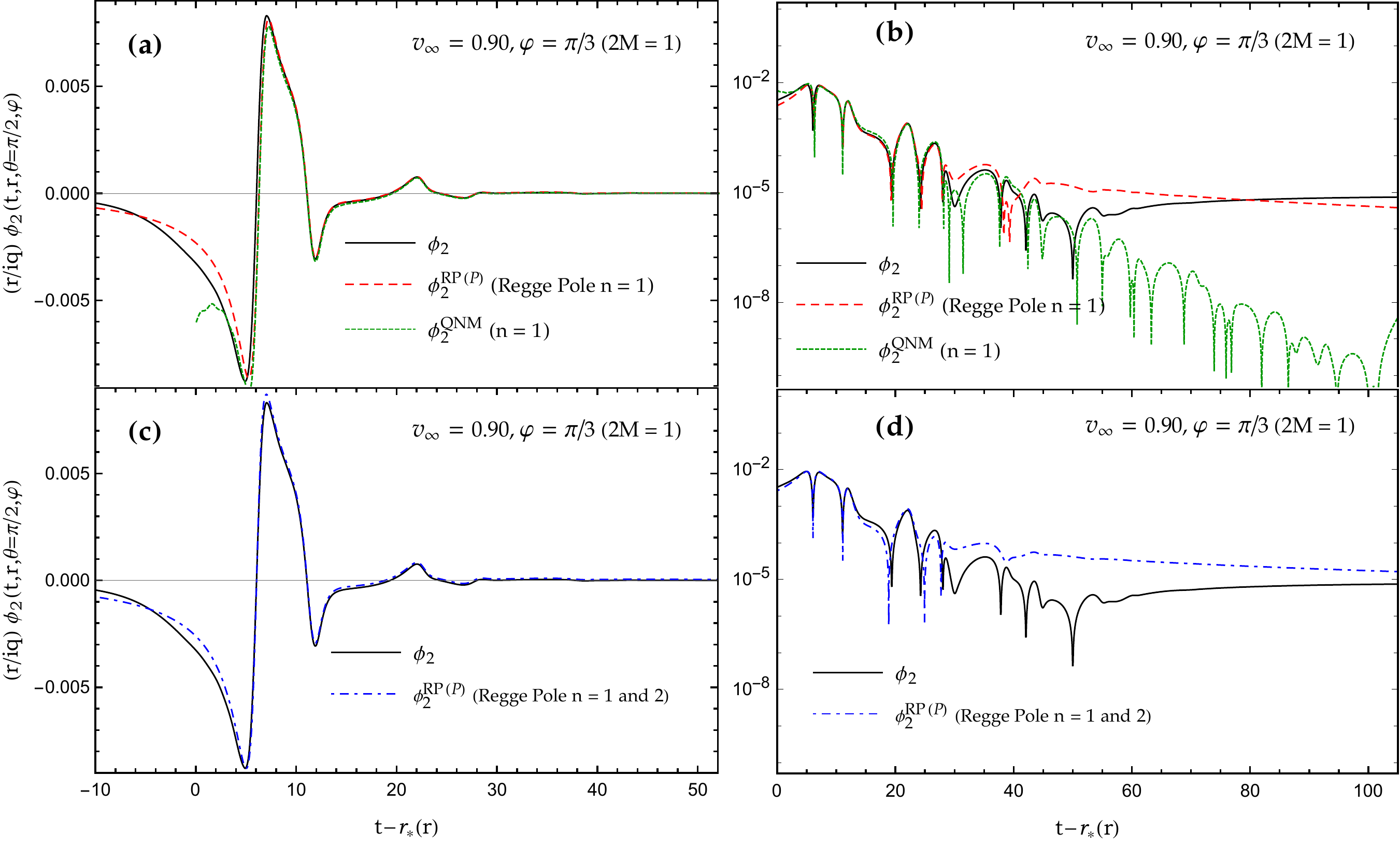}
\caption{\label{P_Exact_QNM_CAM_pis3_v090} The Maxwell scalar $\phi_2$ and its Regge pole approximation $\phi^{\text{\tiny{RP}} \, \textit{\tiny{(P)}}}_2$ for $v_\infty =0.90$ ($\gamma \approx 2.29$) and $\varphi=\pi/3$. (a) The Regge pole approximation constructed from only one Regge pole is in very good agreement with the Maxwell scalar $\phi_2$ constructed by summing over the first fifteen partial waves. The associated quasinormal response $\phi^\text{\tiny{QNM}}_2$  obtained by summing over the $(\ell, n)$ QNMs with $n=1$ and $\ell=1, \dots, 15$ is also displayed. At intermediate timescales, it matches very well the Regge pole approximation. (b) Semilog graph corresponding to (a) and showing that the Regge pole approximation describes very well the ringdown, correctly the pre-ringdown phase and roughly approximates the waveform tail. (c) and (d) Taking into account an additional Regge pole does not improve the Regge pole approximation.}
\end{figure*}

\begin{figure*}
\centering
 \includegraphics[scale=0.49]{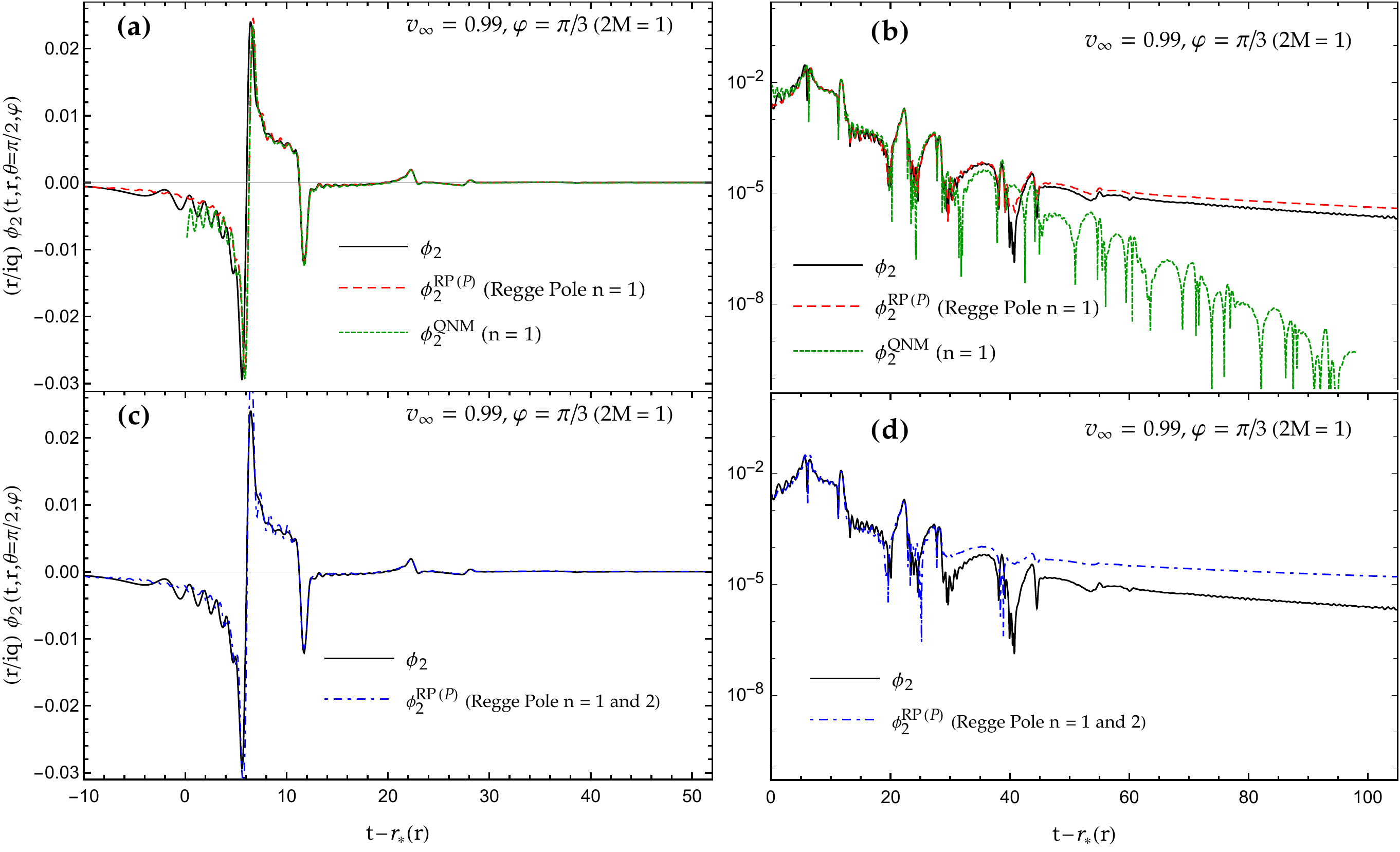}
\caption{\label{P_Exact_QNM_CAM_pis3_v099} The Maxwell scalar $\phi_2$ and its Regge pole approximation $\phi^{\text{\tiny{RP}} \, \textit{\tiny{(P)}}}_2$ for $v_\infty =0.99$ ($\gamma \approx 7.09$) and $\varphi=\pi/3$. (a) The Regge pole approximation constructed from only one Regge pole is in very good (and even impressive) agreement with the Maxwell scalar $\phi_2$ constructed by summing over the first nineteen partial waves. The associated quasinormal response $\phi^\text{\tiny{QNM}}_2$  obtained by summing over the $(\ell, n)$ QNMs with $n=1$ and $\ell=1, \dots, 19$ is also displayed. At intermediate timescales, it matches very well the Regge pole approximation. (b) Semilog graph corresponding to (a) and showing that the Regge pole approximation describes very well the whole signal. (c) and (d) Taking into account an additional Regge pole does not improve the Regge pole approximation.}
\end{figure*}

\begin{figure*}
\centering
 \includegraphics[scale=0.49]{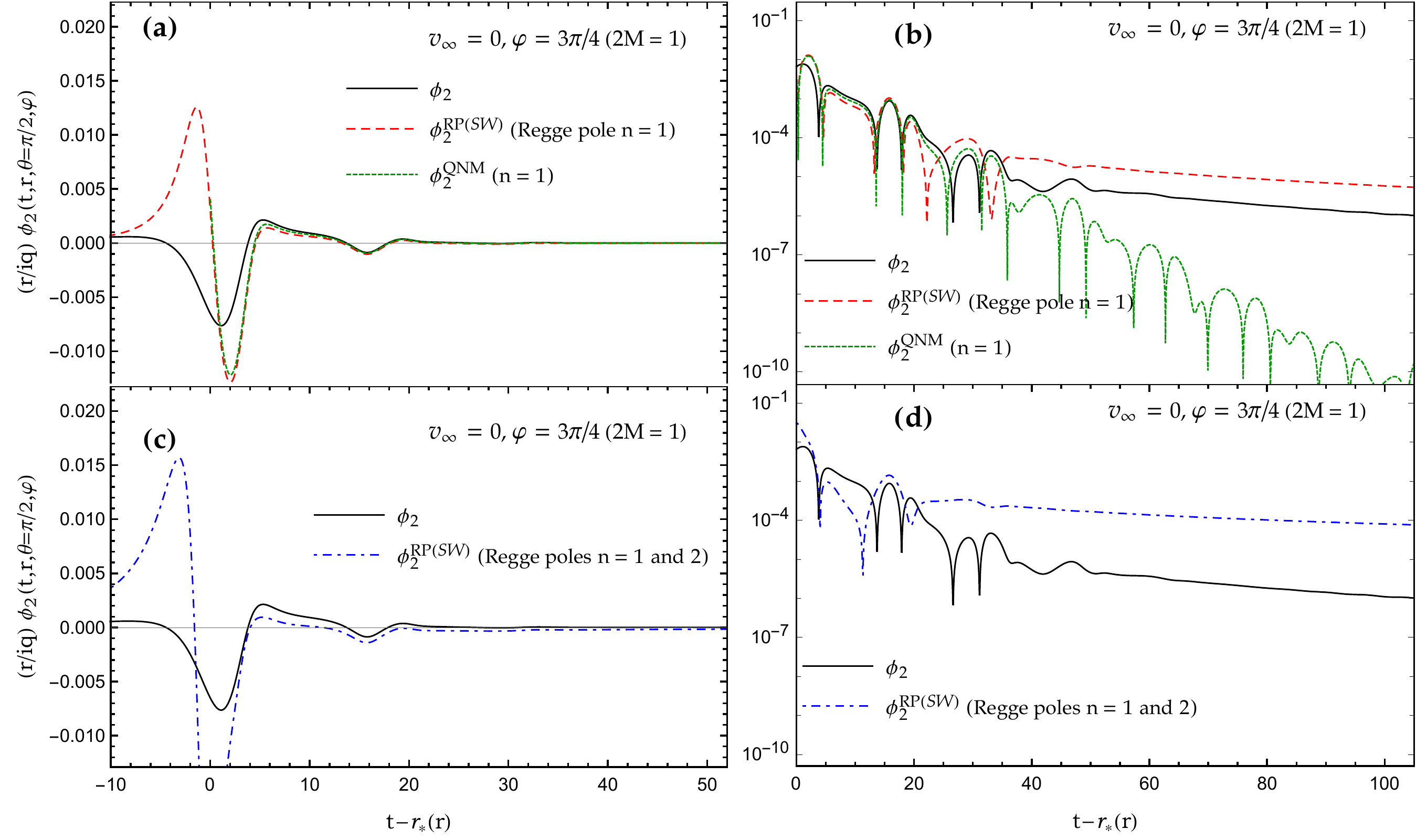}
\caption{\label{SW_Exact_QNM_CAM_3pis4_v0} The Maxwell scalar $\phi_2$ and its Regge pole approximation $\phi^{\text{\tiny{RP}} \, \textit{\tiny{(SW)}}}_2$ for $v_\infty =0$ ($\gamma = 1$) and $\varphi=3\pi/4$. (a) and (b) The pre-ringdown phase of the Maxwell scalar $\phi_2$ constructed by summing over the first thirteen partial waves is not described by the Regge pole approximation constructed from only one Regge pole. However, this approximation matches a large part of the ringdown and roughly approximates the waveform tail. The quasinormal response $\phi^\text{\tiny{QNM}}_2$  obtained by summing over the $(\ell, n)$ QNMs with $n=1$ and $\ell=1, \dots, 13$ is also displayed. At intermediate timescales, it matches correctly the Regge pole approximation. (c) and (d) Taking into account an additional Regge pole does not improve the Regge pole approximation.}
\end{figure*}

\begin{figure*}
\centering
 \includegraphics[scale=0.49]{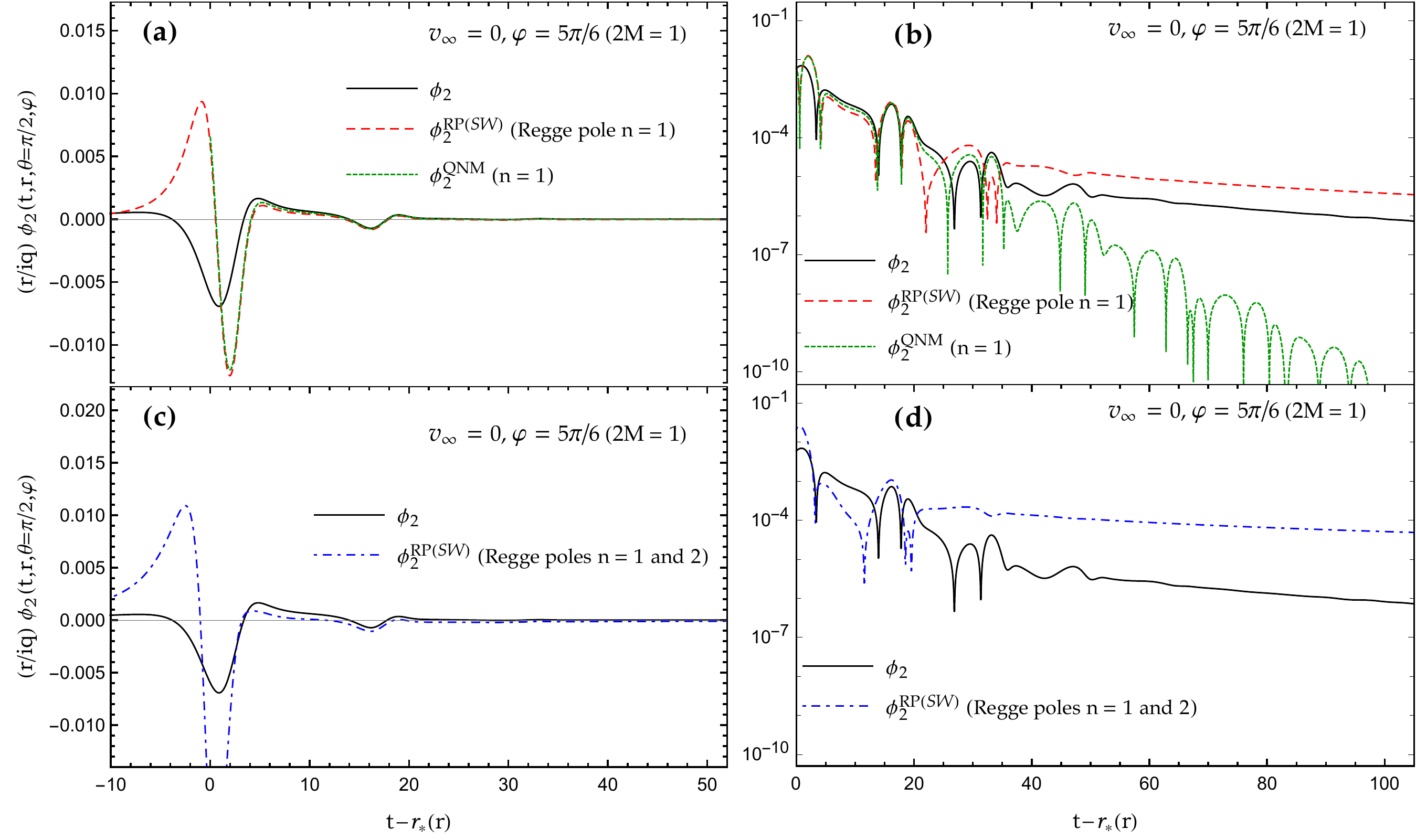}
\caption{\label{SW_Exact_QNM_CAM_5pis6_v0} The Maxwell scalar $\phi_2$ and its Regge pole approximation $\phi^{\text{\tiny{RP}} \, \textit{\tiny{(SW)}}}_2$ for $v_\infty =0$ ($\gamma=1$) and $\varphi=5\pi/6$. (a) and (b) The pre-ringdown phase of the Weyl scalar $\phi_2$ constructed by summing over the first thirteen partial waves is not described by the Regge pole approximation constructed from only one Regge pole. By contrast, this approximation matches correctly a large part of the ringdown and roughly approximates the waveform tail. The quasinormal response $\phi^\text{\tiny{QNM}}_2$ obtained by summing over the $(\ell, n)$ QNMs with $n=1$ and $\ell=1, \dots, 13$ is also displayed. At intermediate timescales, it matches correctly the Regge pole approximation. (c) and (d) Taking into account an additional Regge pole does not improve the Regge pole approximation.}
\end{figure*}

\begin{figure*}
\centering
 \includegraphics[scale=0.49]{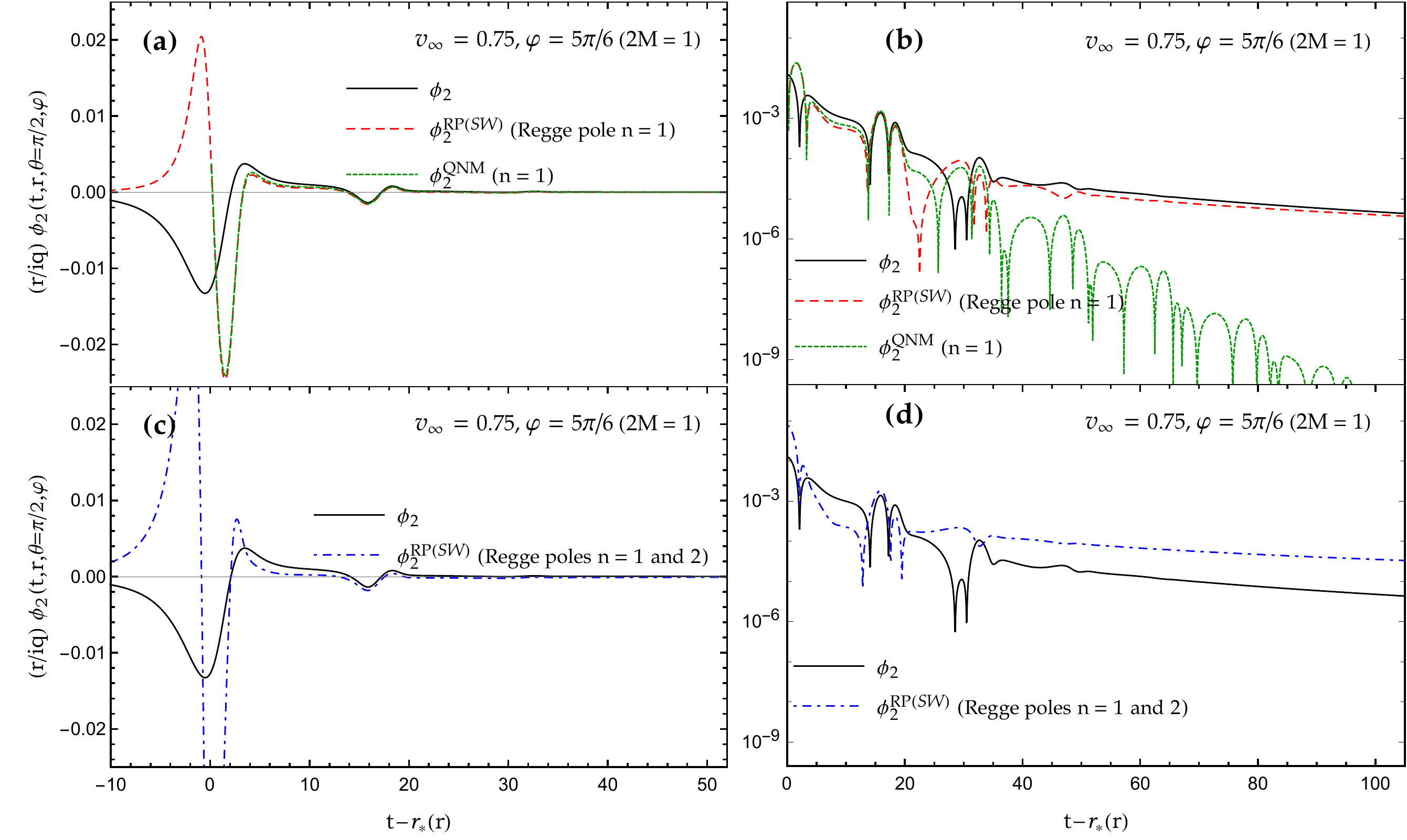}
\caption{\label{SW_Exact_QNM_CAM_5pis6_v075} The Maxwell scalar $\phi_2$ and its Regge pole approximation $\phi^{\text{\tiny{RP}} \, \textit{\tiny{(SW)}}}_2$ for $v_\infty =0.75$ ($\gamma \approx 1.51$) and $\varphi=5\pi/6$. (a) and (b) The pre-ringdown phase of the Maxwell scalar $\phi_2$ constructed by summing over the first fifteen partial waves is not described by the Regge pole approximation constructed from only one Regge pole. By contrast, this approximation matches correctly a large part of the ringdown and the waveform tail. The quasinormal response $\phi^\text{\tiny{QNM}}_2$ obtained by summing over the $(\ell, n)$ QNMs with $n=1$ and $\ell=1, \dots, 15$ is also displayed. At intermediate timescales, it matches very well the Regge pole approximation. (c) and (d) Taking into account an additional Regge pole does not improve the Regge pole approximation.}
\end{figure*}

 \begin{figure*}
\centering
 \includegraphics[scale=0.49]{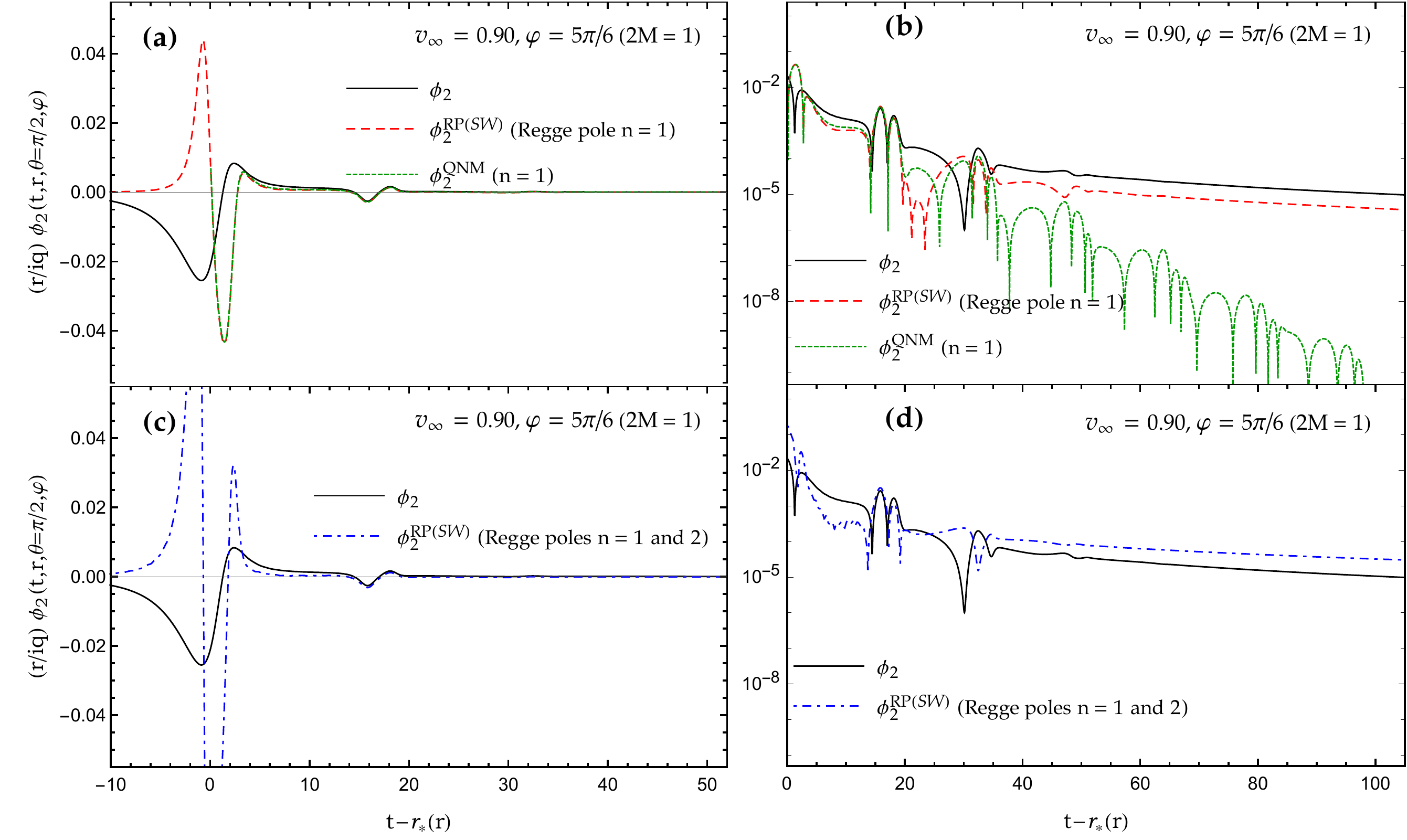}
\caption{\label{SW_Exact_QNM_CAM_5pis6_v090} The Maxwell scalar $\phi_2$ and its Regge pole approximation $\phi^{\text{\tiny{RP}} \, \textit{\tiny{(SW)}}}_2$ for $v_\infty =0.75$ ($\gamma \approx 2.29$) and $\varphi=5\pi/6$. (a) and (b) The pre-ringdown phase of the Maxwell scalar $\phi_2$ constructed by summing over the first fifteen partial waves is not described by the Regge pole approximation constructed from only one Regge pole. By contrast, this approximation matches correctly a large part of the ringdown and roughly approximates the waveform tail. The quasinormal response $\phi^\text{\tiny{QNM}}_2$ obtained by summing over the $(\ell, n)$ QNMs with $n=1$ and $\ell=1, \dots, 15$ is also displayed. At intermediate timescales, it matches very well the Regge pole approximation. (c) and (d) Taking into account an additional Regge pole does not improve the Regge pole approximation.}
\end{figure*}

 \begin{figure*}
\centering
 \includegraphics[scale=0.49]{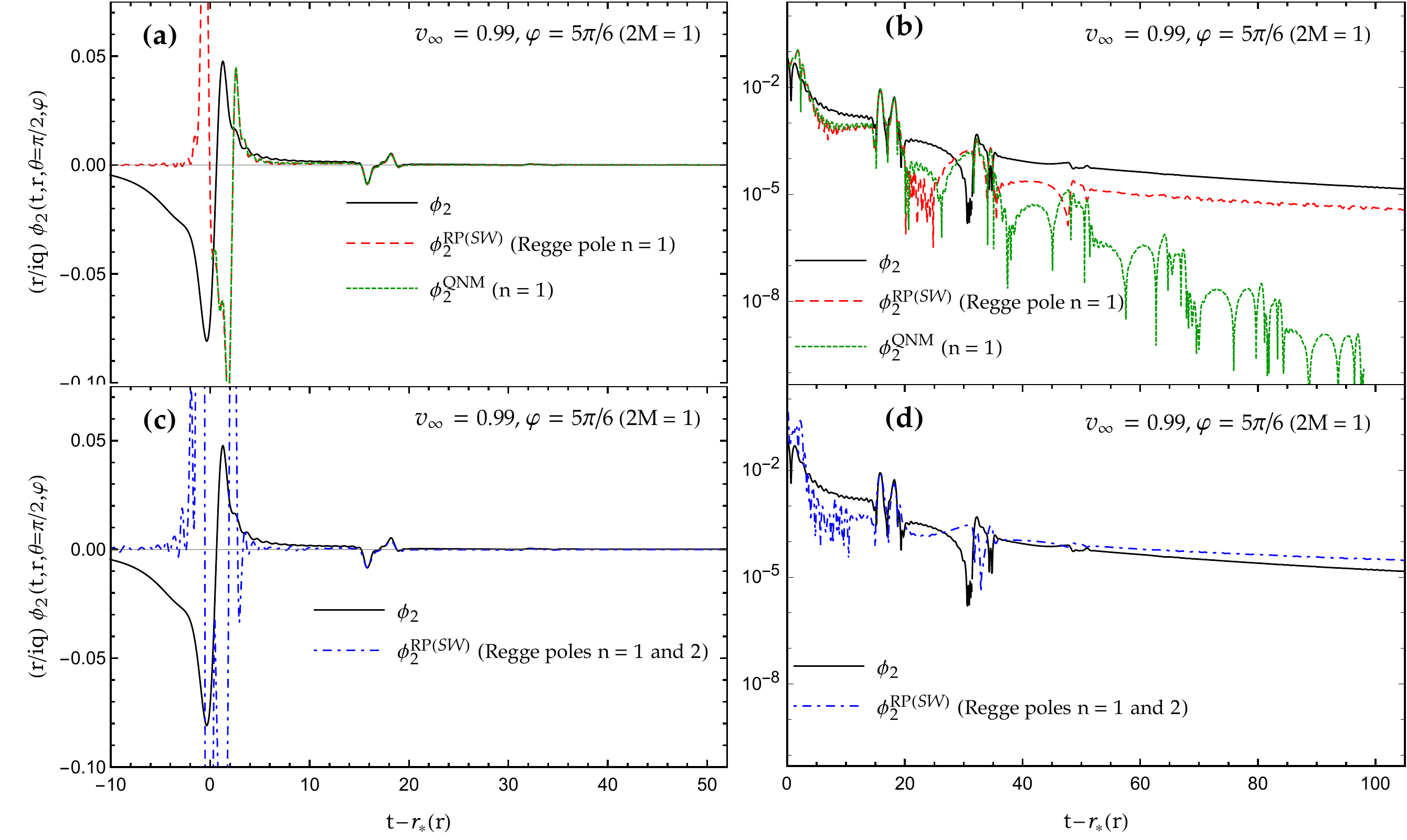}
\caption{\label{SW_Exact_QNM_CAM_5pis6_v099} The Maxwell scalar $\phi_2$ and its Regge pole approximation $\phi^{\text{\tiny{RP}} \, \textit{\tiny{(SW)}}}_2$ for $v_\infty =0.75$ ($\gamma \approx 7.09$) and $\varphi=5\pi/6$. (a) and (b) The pre-ringdown phase of the Maxwell scalar $\phi_2$ constructed by summing over the first nineteen partial waves is not described by the Regge pole approximation constructed from only one Regge pole. This approximation roughly matches the ringdown and the waveform tail. The quasinormal response $\phi^\text{\tiny{QNM}}_2$ obtained by summing over the $(\ell, n)$ QNMs with $n=1$ and $\ell=1, \dots, 19$ is also displayed. At intermediate timescales, it matches very well the Regge pole approximation. (c) and (d) Taking into account an additional Regge pole does not improve the Regge pole approximation.}
\end{figure*}

\subsection{Results and comments}
\label{SecIVb}

We have compared the multipolar waveform $\phi_2$ and the associated quasinormal ringdown with the Regge pole approximations $\phi^{{\text{\tiny{RP}}}\,\textit{\tiny{(P)}}}_2$ in Figs.~\ref{P_Exact_QNM_CAM_pis6_v0}--\ref{P_Exact_QNM_CAM_pis3_v099} and with the Regge pole approximation $\phi^{\text{\tiny{RP}} \, \textit{\tiny{(SW)}}}_2$ in Figs.~\ref{SW_Exact_QNM_CAM_3pis4_v0}--\ref{SW_Exact_QNM_CAM_5pis6_v099}. This has been done for various values of the angle $\varphi \in [0, \pi]$ excluding the cases $\varphi=0$ and $\varphi=\pi$ for which the Maxwell scalar $\phi_2$ vanishes. More precisely, we have considered the case of (i) a particle initially at rest at infinity [$v_\infty = 0$ ($\gamma=1$)], (ii) a particle projected with a relativistic velocity at infinity [we have considered the configurations $v_\infty = 0.75$ ($\gamma\approx 1.51$) and $v_\infty = 0.90$ ($\gamma\approx 2.29$)], and (iii) a particle projected with an ultra-relativistic velocity at infinity [$v_\infty = 0.99$ ($\gamma \approx 7.09$)]. It should be specified that, in order to obtain numerically stable results, the number of partial modes to include in the sum (\ref{phi2_ExpressionDef}) strongly depends on the initial velocity of the particle: the sum over $\ell$ has been truncated at $\ell=13$ for $v_\infty =0$, at $\ell=15$ for $v_\infty =0.75$ and $v_\infty =0.90$, and at $\ell= 19$ for $v_\infty =0.99$. It should be noted that the terminology we used in Ref.~\cite{Folacci:2018sef} to describe the different parts of the multipolar waveform $\Psi_4$ is also adopted for the waveform $\phi_2$: we shall thus designate by ``pre-ringdown phase'' the early time response of the BH and, as usual, we shall refer to the ringdown phase and to the tail of the signal for those parts of the waveform corresponding respectively to intermediate timescales and to very late times.

In Figs.~\ref{P_Exact_QNM_CAM_pis6_v0}--\ref{P_Exact_QNM_CAM_3pis4_v0}, we have compared the multipolar waveform $\phi_2$ generated by a particle initially at rest at infinity with its Regge pole approximation  $\phi^{\text{\tiny{RP}} \, \textit{\tiny{(P)}}}_2$ obtained from the Poisson summation formula. In Figs.~\ref{P_Exact_QNM_CAM_pis6_v0}--\ref{P_Exact_QNM_CAM_pis2_v0}, for $\varphi = \pi/6, \pi/3$ and $\pi/2$, we can observe that the Regge pole approximation constructed from only one Regge pole is in good or very good agreement with the exact waveform, and that an additional Regge pole does not really improve this approximation. More precisely, it is interesting to note that the Regge pole approximation matches the ringdown, describes correctly the pre-ringdown phase and roughly the waveform tail. It is moreover important to note that it provides a description of the ringdown that does not necessitate determining a starting time, in contrast to the ringdown waveform constructed from the QNMs which is exponentially divergent as $t$ decreases. In Fig.~\ref{P_Exact_QNM_CAM_3pis4_v0}, for $\varphi = 3\pi/4$, we can observe that the Regge pole approximation is no longer so interesting. Indeed, it only roughly describes the BH response. Here, it should be recall that the Regge pole approximation $\phi^{\text{\tiny{RP}}\,\textit{\tiny {(P)}}}_2 $ diverges for $\varphi \to \pi$ and, as a consequence, for $\varphi=3\pi/4$ (i.e., for a value of $\varphi$ rather close to $\pi$), it would be necessary to consider the background integral contribution $\phi^{\text{\tiny{B}} \, \textit{\tiny{(P)}}}_2$ given by Eq.~(\ref{CAM_phi2_ExpressionDef_P_Background}) to correctly describe the multipolar waveform $\phi_2$.

In Figs.~\ref{P_Exact_QNM_CAM_pis6_v075}--\ref{P_Exact_QNM_CAM_pis3_v099}, we have compared, for $\varphi= \pi/6 $ and $\pi/3$, the multipolar waveform $\phi_2$ generated by a particle projected with a relativistic or an ultra-relativistic velocity at infinity with the Regge pole approximation $\phi^{\text{\tiny{RP}}\,\textit{\tiny{(P)}}}_2$ obtained from the Poisson summation formula. We can observe that the whole signal is impressively described by the Regge pole approximation constructed from only one Regge pole and that this approximation is even more efficient in the ultra-relativistic context.

In Fig.~\ref{SW_Exact_QNM_CAM_3pis4_v0}, for $\varphi = 3 \pi/4$, we have compared the multipolar waveform $\phi_2$ generated by a particle initially at rest at infinity with the Regge pole approximation $\phi^{\text{\tiny{RP}}\,\textit{\tiny{(SW)}}}_2 $ obtained from the Sommerfeld-Watson transformation. We recall that, while $\phi^{\text{\tiny{RP}}\,\textit{\tiny{(P)}}}_2$ constructed from the Poisson summation formula diverges in the limit $\varphi \to \pi $, the Regge pole approximation $\phi^{\text{\tiny{RP}}\,\textit{\tiny{(SW)}}}_2 $ is regular in the same limit (it only diverges for $\varphi \to 0$). As a consequence, the latter approximation should provide better results than the former one for $\varphi$ close to $\pi$. By comparing Fig.~\ref{SW_Exact_QNM_CAM_3pis4_v0} with Fig.~\ref{P_Exact_QNM_CAM_3pis4_v0}, we can see that this seems to be the case if we focus on the ringdown phase of the waveform but that the pre-ringdown phase is not described at all. In fact, here, to correctly describe the waveform $\phi_2$ we should take into account the background integral contribution $\phi^{\text{\tiny{B}}\, \textit{\tiny{(SW)}}}_2$ given by Eq.~(\ref{CAM_phi2_ExpressionDef_SW_Background}).

In Figs.~\ref{SW_Exact_QNM_CAM_5pis6_v0}--\ref{SW_Exact_QNM_CAM_5pis6_v099}, for $ \varphi = 5\pi/6 $, we have displayed the multipolar waveform $\phi_2$ generated by a particle initially at rest at infinity and by a particle projected with a relativistic or an ultra-relativistic velocity, and we have compared it with the Regge pole approximation $\Psi^{\text{\tiny {RP}}\,\textit{\tiny {(SW)}}}_4$ obtained from the transformation of Sommerfeld-Watson. Here again, the Regge pole approximation constructed from a single Regge pole does not describe the pre-ringdown phase of the Maxwell scalar $\phi_2$, but it matches a large part of the ringdown phase and approximates the tail rather correctly.

\section{Electromagnetic energy spectrum  $dE/d\omega$ and its CAM representation}
\label{SecV}

In this section, we shall focus on the electromagnetic energy spectrum $dE/d\omega$ observed at infinity which is generated by the charged particle falling radially into the Schwarzschild BH. We shall provide its CAM representation from the Poisson summation formula and Cauchy's theorem and discuss the interest of this representation and of the corresponding Regge pole approximation.

\subsection{Total energy radiated by the particle and associated electromagnetic energy spectrum}
\label{SecV_a}

The electromagnetic power ${\cal P}$ radiated at spatial infinity by the charged particle, i.e., the rate $dE/dt$ at which the electromagnetic field generated by this particle carries energy to infinity, can be obtained as the flux of the Poynting vector ${\mathbf R}$ across a spherical surface $S(r)$ with radius $r \to \infty$: we have
\begin{eqnarray}\label{dEdt}
 {\cal P}= \frac{dE}{dt} = \lim_{r \to \infty} \int_{S(r)} {\mathbf R} \cdot  {\mathbf {dS}}
\end{eqnarray}
with ${\mathbf R}={\mathbf E} \wedge {\mathbf B}$ and $\mathbf {dS} = r^2 \sin\theta \, d\theta \, d\varphi \, {\bf {\hat{e}}_r}$. By using Eqs.~(\ref{ChampE_even}), (\ref{Relations_ChampB_even_odd}) and (\ref{TF_psi_bis_a}) as well as the orthonormalization relation (\ref{HSV_even_Norm}) for the vector spherical harmonics and the addition theorem for scalar spherical harmonics (\ref{ThAd_HS}), we obtain
\begin{subequations}\label{dEdt}
\begin{eqnarray}\label{dEdt_a}
  \frac{dE}{dt}(t) &=&\frac{1}{4\pi} \sum_{\ell m} \frac{2 \ell +1 }{\ell(\ell+1)}\, \Big{|} \partial_t \psi_{\ell} (t,r \to +\infty) \Big{|}^{2}
\end{eqnarray}
or, more explicitly, by using Eqs.~(\ref{TF_psi_bis_b}) and (\ref{Partial_Response_c}),
\begin{eqnarray}\label{dEdt_b}
 & &  \frac{dE}{dt}(t) =\frac{q^2}{4\pi}\sum_{\ell = 1}^{+\infty} (2\ell+1)\ell(\ell+1) \nonumber \\
 & & \qquad \quad \times \Bigg{|} \frac{1}{\sqrt{2\pi}}\int_{-\infty}^{+\infty}d\omega\, \frac{e^{-i  \omega[t-r_\ast(r)]}}{2i\omega A_{\ell}^{(-)}(\omega)}\widetilde{K}[\ell,\omega]\Bigg{|}^2
\end{eqnarray}
\end{subequations}
with $r \to +\infty$.

The previous result provides, by integration over $t$, the total energy $E$ radiated by the charged particle during its fall in the BH. We have
\begin{subequations}
\begin{eqnarray}\label{Etot_INTt}
& & E =\int_{-\infty}^{+\infty} dt \, \frac{dE}{dt}(t) \label{Etot_INTt_a}\\
& & \phantom{E} = \frac{q^2}{4\pi}\sum_{\ell = 1}^{+\infty} (2\ell+1)\ell(\ell+1) \nonumber \\
& & \qquad \times \int_{-\infty}^{+\infty} dt \, \Bigg{|} \frac{1}{\sqrt{2\pi}}\int_{-\infty}^{+\infty}d\omega\, \frac{e^{-i  \omega t}}{2i\omega A_{\ell}^{(-)}(\omega)}\widetilde{K}[\ell,\omega]\Bigg{|}^2  \nonumber \\
& & \label{Etot_INTt_b}
\end{eqnarray}
\end{subequations}
[note that the dependence in $r$ now disappears due to the change of variable $t \to t + r_\ast(r)$]. We can obtain an alternative expression for the total energy $E$ by applying the Parseval-Plancherel theorem to Eq.~(\ref{Etot_INTt_b}). This gives immediately
\begin{eqnarray}\label{Etot_PP}
& & E = \frac{q^2}{4\pi}\sum_{\ell = 1}^{+\infty} (2\ell+1)\ell(\ell+1) \int_{-\infty}^{+\infty} d\omega \, \Bigg{|} \frac{\widetilde{K}[\ell,\omega]}{2i\omega A_{\ell}^{(-)}(\omega)}\Bigg{|}^2. \nonumber \\
& &
\end{eqnarray}
This new form of $E$ permits us to derive the expression of the (total) electromagnetic energy spectrum $dE/d\omega$ radiated by the particle. Indeed, from a physical point of view, it is defined for $\omega \ge 0$ and satisfy
\begin{equation}\label{dEdw_w_def}
E = \int_0^{+\infty} d\omega \,  \frac{dE}{d\omega} (\omega).
\end{equation}
Then, by using Eq.~(\ref{Sym_om_d}) in Eq.~(\ref{Etot_PP}), we obtain
\begin{eqnarray}\label{Etot_PP_bis}
& & E = \frac{q^2}{2\pi}\sum_{\ell = 1}^{+\infty} (2\ell+1)\ell(\ell+1) \int_{0}^{+\infty} d\omega \, \Bigg{|} \frac{\widetilde{K}[\ell,\omega]}{2i\omega A_{\ell}^{(-)}(\omega)}\Bigg{|}^2 \nonumber \\
& &
\end{eqnarray}
and by comparing Eq.~(\ref{Etot_PP_bis}) with Eq.~(\ref{dEdw_w_def}) we have
\begin{subequations}\label{dEdw_bis}
\begin{equation}\label{dEdw_bis_a}
  \frac{dE}{d\omega} (\omega) = \sum_{\ell=1}^{+\infty} \frac{dE_\ell}{d\omega}(\omega)
\end{equation}
where
\begin{equation}\label{dEdw_bis_b}
\frac{dE_\ell}{d\omega}(\omega) = \frac{q^2}{8\pi \omega^2} \times (2\ell+1)\ell(\ell+1)\, \Gamma_\ell(\omega) \, \Big{|}\widetilde{K}[\ell,\omega]\Big{|}^2
\end{equation}
\end{subequations}
denotes the partial energy spectrum corresponding to the $\ell$th mode. It is very important to note that in Eq.~(\ref{dEdw_bis_b}), we have chosen to introduce explicitly the greybody factors
\begin{equation}\label{Greybody_factors}
  \Gamma_{\ell}(\omega)= \frac{1}{\big{|}A_{\ell}^{(-)}(\omega) \big{|}^2}
\end{equation}
of the Schwarzschild BH corresponding to the electromagnetic field. It is worth pointing out that we can write $E$ given by Eqs.~(\ref{dEdw_w_def}) and (\ref{dEdw_bis}) in the form
\begin{subequations}\label{Etot}
\begin{equation}\label{Etot_a}
  E = \sum_{\ell=1}^{+\infty} E_\ell
\end{equation}
where
\begin{equation}\label{Etot_b}
E_\ell= \int_0^{+\infty} d\omega \, \frac{dE_\ell}{d\omega}(\omega)
\end{equation}
\end{subequations}
denotes the partial energy radiated in the $\ell$th mode.

Finally, it is important to note that Eq.~(\ref{dEdw_bis}) can also be written as
\begin{equation}\label{dEdw_bis2}
  \frac{dE}{d\omega}(\omega) =\frac{q^2}{8\pi \omega^2}\sum_{\ell = 0}^{+\infty} (2\ell+1)\ell(\ell+1)\, \Gamma_\ell(\omega) \, \Big{|}\widetilde{K}[\ell,\omega]\Big{|}^2.
\end{equation}
Indeed, here again, as in Sec.~\ref{SecIIe}, it is possible to start at $\ell=0$ the discrete sum over $\ell$ due to the relations (\ref{Formally_ell0}). In the next subsection, we shall take Eq.~(\ref{dEdw_bis2}) as a starting point because it will permit us to use the Poisson summation formula in its standard form.

\subsection{CAM representation based on the Poisson summation formula}
\label{SecV_b}

In order to start the CAM machinery permitting us to derive a CAM representation of the electromagnetic energy spectrum $dE / d\omega$, it is necessary to replace in Eq.~(\ref{dEdw_bis2}) the angular momentum $\ell \in \mathbb{N}$ by the angular momentum $\lambda = \ell +1/2 \in \mathbb{C}$ and therefore to have at our disposal the analytic extensions in the complex $\lambda$ plane of all the functions of $\ell$ appearing in Eq.~(\ref{dEdw_bis2}). In fact, in Sec.~\ref{SecIIIa}, we have already discussed the construction of the analytic extensions of $A^{(\pm)}_{\ell} (\omega)$ and $\widetilde{K}[\ell,\omega]$. It should be however noted that, here, the situation is a little bit more complicated: indeed, we need the analytic extensions of $\Gamma_{\ell}(\omega)= 1/ \big{|}A_{\ell}^{(-)}(\omega) \big{|}^2$ and $\big{|}\widetilde{K}[\ell,\omega]\big{|}^2$. Fortunately, in Sec.~II of Ref.~\cite{Decanini:2011xi} where the absorption problem for a massless scalar field propagating in a Schwarzschild BH has been considered, the analytic extension $\Gamma_{\lambda-1/2}(\omega)$ of the greybody factor $\Gamma_\ell(\omega)$ has been discussed. Here, we shall adopt the same prescription, i.e., we shall assume that
\begin{equation}\label{Gamma_Analytic_Extension}
  \Gamma_{\lambda-1/2}(\omega) =\frac{1}{A_{\lambda-1/2}^{(-)}(\omega) \, {[A^{(-)}_{\lambda^\ast -1/2} (\omega) ]}^\ast}.
\end{equation}
We recall that this particular extension permits us to work with an even function of $\lambda$ which is purely real. (For more details concerning the  properties of the greybody factor $\Gamma_{\lambda-1/2}(\omega)$, we refer to Sec.~II of Ref.~\cite{Decanini:2011xi}.) Furthermore, we shall adopt an analogous prescription for the analytic extension of $\big{|}\widetilde{K}[\ell,\omega]\big{|}^2$ by considering that it is given by $\widetilde{K}[\lambda-1/2,\omega] \, {[ \widetilde{K}[\lambda^\ast -1/2,\omega] ]}^\ast$.

In order to derive a CAM representation of the electromagnetic energy spectrum $dE / d\omega$, the use of CAM techniques requires in addition the determination of the singularities of the analytic extensions in the complex $\lambda$ plane of all the functions of $\ell$ appearing in Eq.~(\ref{dEdw_bis2}). Here, the only singularities to consider are the simple poles of the greybody factor $\Gamma_{\lambda-1/2}(\omega)$. In fact, they have been also studied in Sec.~II of Ref.~\cite{Decanini:2011xi}. Let us just recall that:
\begin{enumerate}[label=(\roman*)]
  \item The singularities of the function $\Gamma_{\lambda-1/2}(\omega)$ are the Regge poles $\lambda_n(\omega)$, i.e., the zeros of the function $A^{(-)}_{\lambda_n(\omega)-1/2} (\omega)$ [see Eq.~(\ref{PR_def_Am})], as well their complex conjugates ${[\lambda_n(\omega)]}^\ast $, i.e., the zeros of the function ${[A^{(-)}_{\lambda^\ast -1/2} (\omega) ]}^\ast$. For $\omega >0$, the Regge poles $\lambda_n(\omega)$ lie in the first and in the third quadrant of the CAM plane, symmetrically distributed with respect to the origin $O$ of this plane and, as a consequence, the Regge poles ${[\lambda_n(\omega)]}^\ast $ lie in the second and in the fourth quadrant of this plane.
  \item The residues of the function $\Gamma_{\lambda-1/2}(\omega)$ at the poles $\lambda_n(\omega)$ and ${[\lambda_n(\omega)]}^\ast $ are complex conjugate of each other and we have in particular
\begin{eqnarray}\label{gamma_residues}
  \gamma_n(\omega) &=&\text{Res}[\Gamma_\lambda-1/2(\omega)]_{\lambda=\lambda_n(\omega)} \nonumber \\
                   &=&  \frac{1}{\left[\left(\frac{d}{d \lambda}A_{\lambda -1/2}^{(-)}(\omega) \right)  {[A^{(-)}_{\lambda^\ast -1/2} (\omega) ]}^\ast\right]_{\lambda=\lambda_n(\omega)}}.
\end{eqnarray}
\end{enumerate}

We have now at our disposal all the ingredients permitting us to obtain a CAM representation of the electromagnetic energy spectrum $dE / d\omega$ by using the Poisson summation formula~\cite{MorseFeshbach1953} as well as Cauchy's residue theorem. In fact, this can be achieved by following, \textit{mutatis mutandis}, the reasoning of Sec.~II of Ref.~\cite{Decanini:2011xi} where a CAM representation of the absorption cross section of the Schwarzschild BH has been derived [we invite the reader to compare Eq.~(3) of Ref.~\cite{Decanini:2011xi} with Eq.~(\ref{dEdw_bis2}) of the present article]. Taking into account the previous considerations concerning the greybody factor $\Gamma_{\lambda-1/2}(\omega)$, its poles and the associated residues, we obtain
\begin{widetext}
\begin{equation}\label{dEdw_CAM_2}
  \frac{dE}{d\omega} (\omega)= \frac{dE}{d\omega}^{\text{\tiny{B,Re}}}(\omega) + \frac{dE}{d\omega}^{\text{\tiny{B,Im}}}(\omega) + \frac{dE}{d\omega}^{\text{\tiny{RP}}}(\omega)
\end{equation}
where
\begin{subequations}\label{dEdw_CAM_Background}
\begin{equation}\label{dEdw_background_integral_Re}
   \frac{dE}{d\omega}^{\text{\tiny{B,Re}}} (\omega)= \frac{q^2}{4\pi \omega^2} \int_{0}^{+\infty} d\lambda\, \lambda(\lambda^2-1/4) \Gamma_{\lambda-1/2}(\omega) \Big{|}\widetilde{K}[\lambda-1/2,\omega]\Big{|}^2
\end{equation}
is a background integral contribution along the real axis,
\begin{equation}\label{dEdw_background_integral_Im}
   \frac{dE}{d\omega}^{\text{\tiny{B,Im}}} (\omega) = -\frac{q^2}{4 \pi \omega^2}\int_{0}^{+i\infty} d\lambda \, \lambda(\lambda^2-1/4) \Gamma_{\lambda-1/2}(\omega) \Big{|}\widetilde{K}[\lambda-1/2,\omega]\Big{|}^2  \frac{e^{i \pi \lambda}}{\cos(\lambda \pi)}
\end{equation}
\end{subequations}
is a  background integral contribution along the imaginary axis and
\begin{eqnarray}\label{dEdw_CAM_PR}
& &  \frac{dE}{d\omega}^{\text{\tiny{RP}}} (\omega) = -\frac{q^2}{2 \omega^2} \operatorname{Re} \left(\sum_{n=1}^{+\infty} \frac{e^{i\pi[\lambda_n(\omega)-1/2]}\lambda_n(\omega) \left(\lambda_n(\omega)^2-1/4\right) \, \gamma_n(\omega)}{\sin[\pi(\lambda_n(\omega)-1/2)]} \right. \nonumber \\
& &  \left. \qquad\qquad\qquad\qquad\qquad \phantom{\sum_{n=1}^{+\infty}}  \times \widetilde{K}[\lambda_n(\omega)-1/2,\omega] \, \left[\widetilde{K}[{[\lambda_n(\omega)]}^\ast-1/2,\omega]\right]^\ast\, \right)
\end{eqnarray}
\end{widetext}
is a sum over the Regge poles lying in the first quadrant of the CAM plane. Of course, Eqs.~\eqref{dEdw_CAM_2}, \eqref{dEdw_CAM_Background} and \eqref{dEdw_CAM_PR} provide an exact CAM representation of the electromagnetic energy spectrum $dE/d\omega$, equivalent to the initial partial wave expansion~\eqref{dEdw_bis}.

\begin{figure*}
\centering
 \includegraphics[scale=0.65]{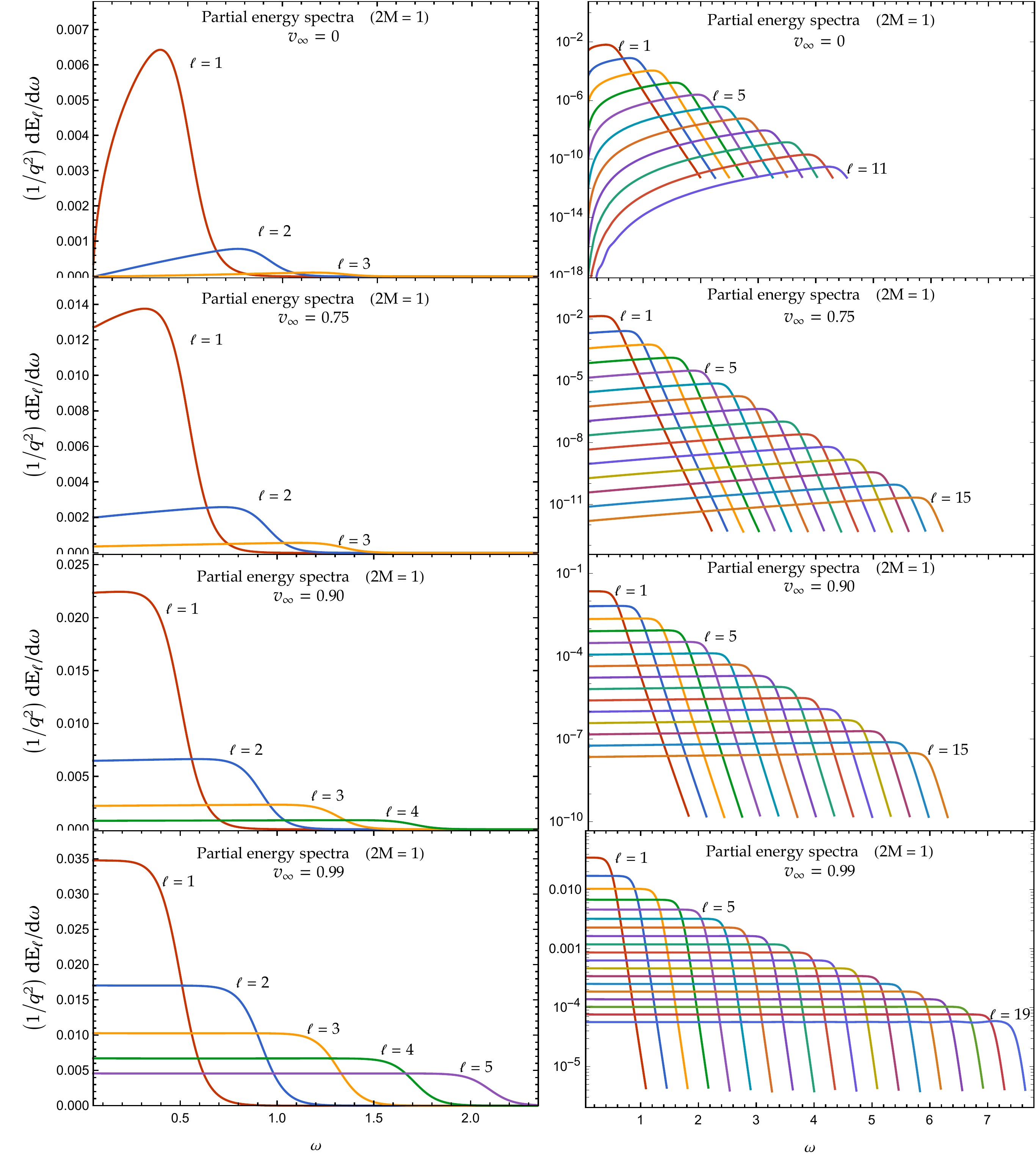}
\caption{\label{dEdw_v_0_075_090_099} The partial electromagnetic energy spectra radiated by a charged falling radially into a Schwarzschild BH. The results are given for $v_\infty =0, 0.75, 0.90$ and $0.99$.}
\end{figure*}

\begin{figure}
\centering
 \includegraphics[scale=0.60]{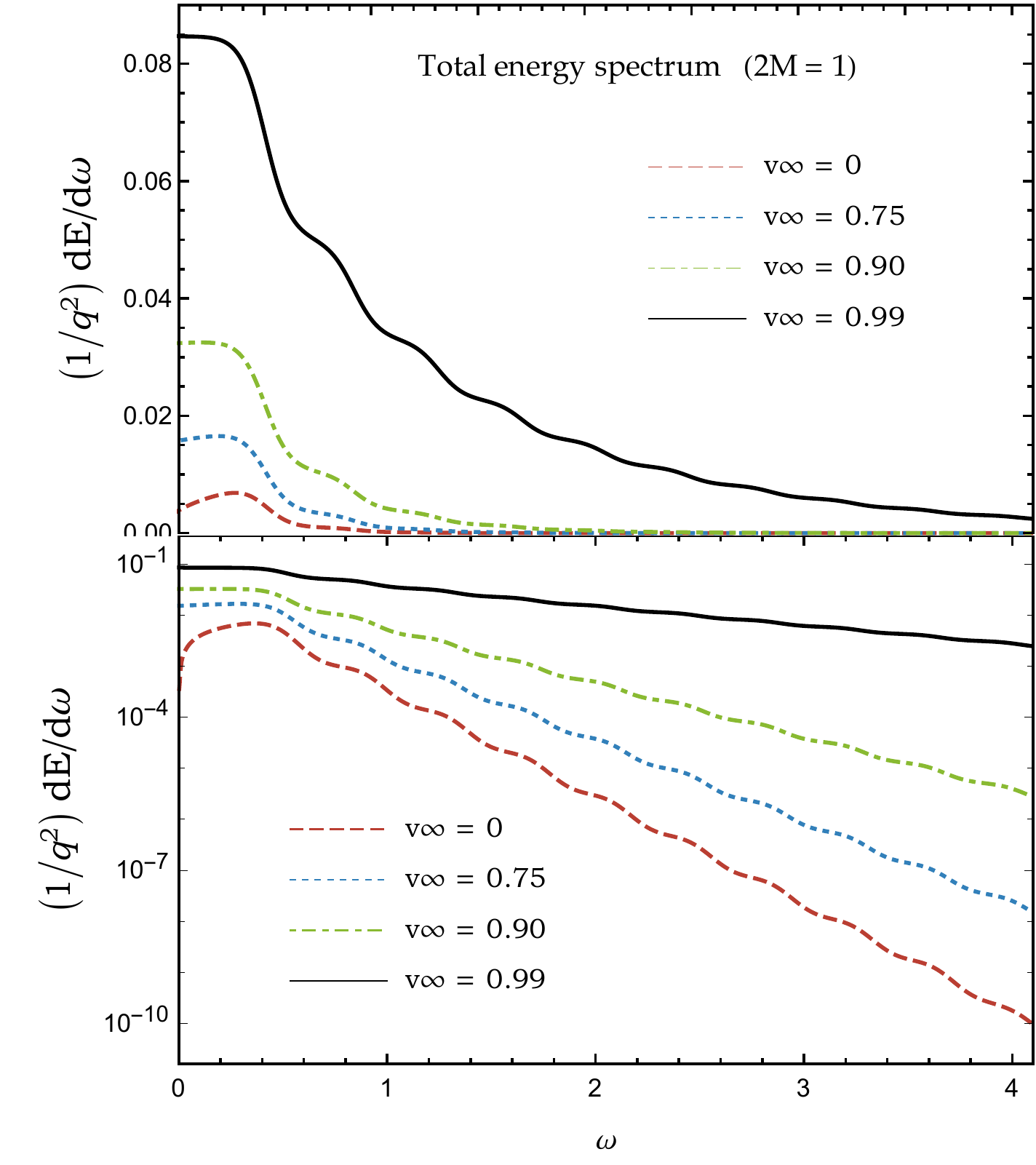}
\caption{\label{dEdw_v_0_075_090_099_tot}  The total electromagnetic energy spectrum radiated by a charged falling radially into a Schwarzschild BH. The results are given for $v_\infty =0, 0.75, 0.90$ and $0.99$.}
\end{figure}

\subsection{Computational methods}
\label{SecV_c}

To construct numerically the electromagnetic energy spectrum~\eqref{dEdw_bis} radiated by a charged particle falling radially into the Schwarzschild BH and its CAM representation~\eqref{dEdw_CAM_2}-\eqref{dEdw_CAM_PR}, we have used the computational methods that have allowed us to obtain numerically the Maxwell scalar $\phi_2$ and its Regge pole approximations in Sec.~\ref{SecIV}. It should be noted that here, we have in addition evaluated the background integral along the real axis (\ref{dEdw_background_integral_Re}) by taking $\lambda \in [0, 25]$ and the background integral along the imaginary axis (\ref{dEdw_background_integral_Im}) by taking  $\lambda \in [0, 6i]$ (due to the term $e^{i \pi \lambda}/\cos[\lambda \pi]$ in the expression of its integrand, this integral converges rapidly).

\begin{figure*}
\centering
 \includegraphics[scale=0.65]{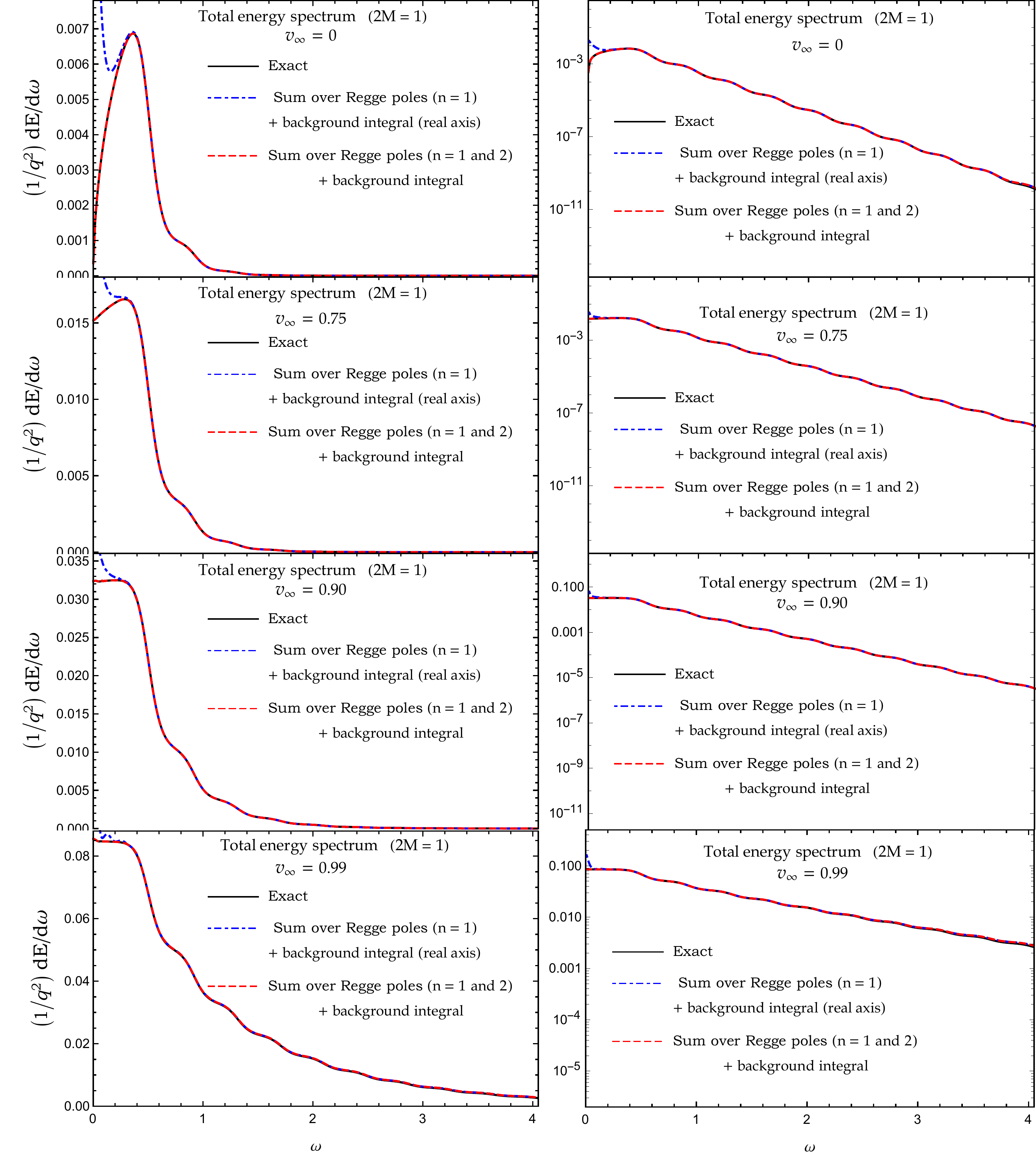}
\caption{\label{dEdw_Exact_vs_CAM_v_0_075_090_099} The electromagnetic energy spectrum radiated by a charged particle falling radially into a Schwarzschild BH compared with its CAM representation. The respective roles of the background integrals (\ref{dEdw_background_integral_Re}) and (\ref{dEdw_background_integral_Im}) and of the Regge pole sum (\ref{dEdw_CAM_PR}) clearly appear.}
\end{figure*}

\begin{figure}
\centering
 \includegraphics[scale=0.60]{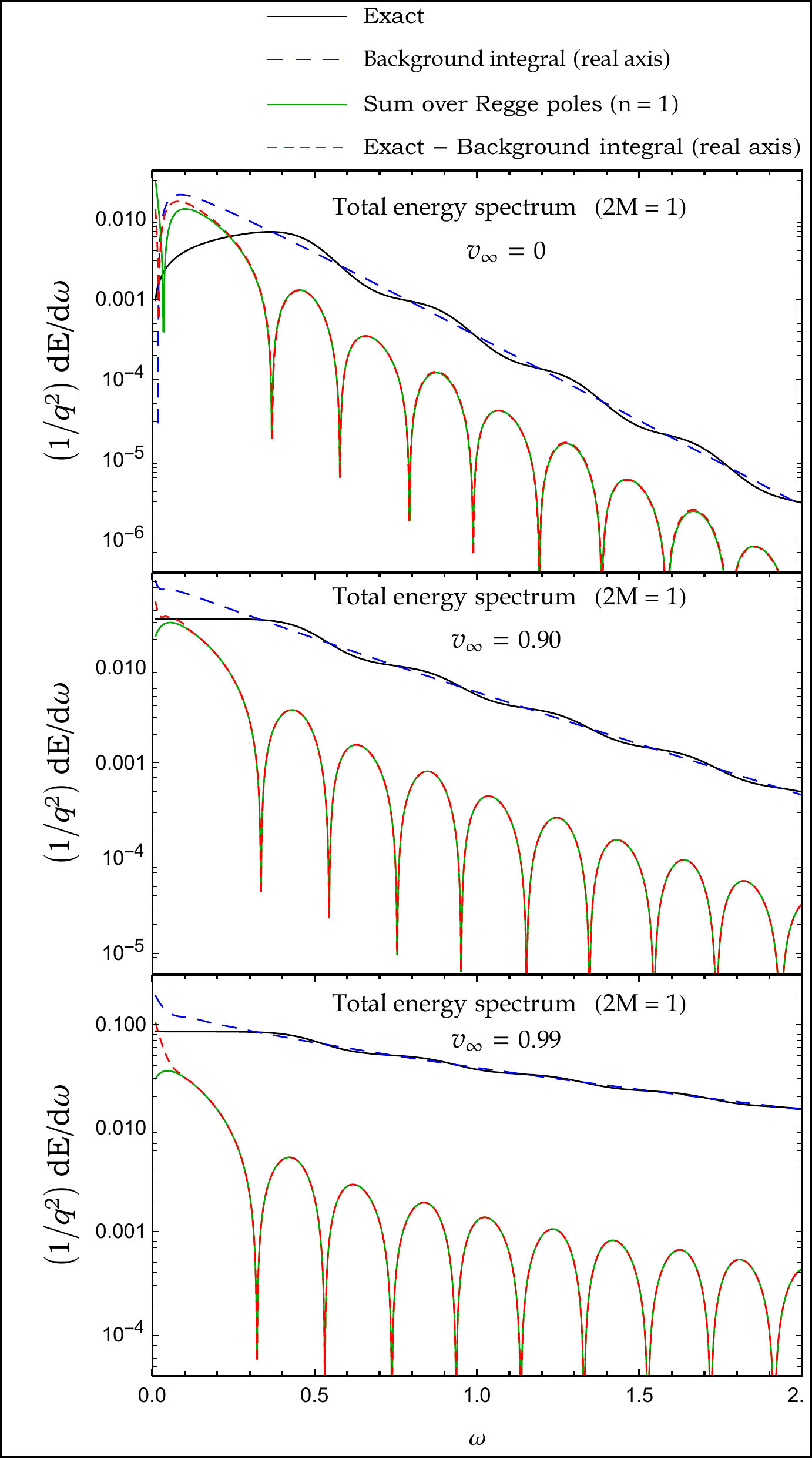}
\caption{\label{dEdw_Exact_vs_CAM_PR_Int_axe_Re_v_0_090_099} The oscillations in the electromagnetic energy spectrum radiated by a charged particle falling radially into a Schwarzschild BH explained by the Regge pole approximation.}
\end{figure}

\subsection{Numerical results and comments}
\label{SecV_d}

We now display and discuss a few results concerning the electromagnetic energy radiated by the charged particle falling radially into a Schwarzschild BH. Here again, as in Sec.~\ref{SecIVb}, we have focused our attention on (i) a particle initially at rest at infinity [$v_\infty =0$ ($\gamma=1$)], (ii) a particle projected with a relativistic velocity at infinity [we have considered the configurations $v_\infty = 0.75$ ($\gamma\approx 1.51$) and $v_\infty = 0.90$ ($\gamma\approx 2.29$)], and (iii) a particle projected with an ultra-relativistic velocity at infinity [$v_\infty = 0.99$ ($\gamma \approx 7.09$)].

In Fig.~\ref{dEdw_v_0_075_090_099}, we have displayed some partial electromagnetic energy spectra $dE_\ell/d\omega$ corresponding to the lowest modes. Our results are in perfect agreement with those already obtained in the literature (see Refs.~\cite{Ruffini:1972uh,Cardoso:2003cn} but note in these articles, the authors used Gaussian units while we consider electromagnetism in the Heaviside system). In Fig.~\ref{dEdw_v_0_075_090_099_tot}, we have displayed the total electromagnetic energy spectrum $dE/d\omega$ for the configurations considered in Fig.~\ref{dEdw_v_0_075_090_099}. It should be noted that, in order to obtain numerically stable results, the number of modes to include in the sum (\ref{dEdw_bis_a}) strongly depends on the initial velocity of the particle. This clearly appears if we examine the ordinate scales used in the semilog graphs of Fig.~\ref{dEdw_v_0_075_090_099}. In fact, we have truncated the sum over $\ell$ at $\ell=10$ for $v_\infty =0$, at $\ell=15$ for $v_\infty =0.75$ and $v_\infty =0.90$, and at $\ell= 20$ for $v_\infty =0.99$.

In Table~\ref{tab:table1}, we have used Eq.~(\ref{Etot}) to compute the total energy $E$ radiated by the charged particle for the values $v_\infty =0, 0.75, 0.90$ and $0.99$ of its velocity at infinity. As expected, $E$ increases with $v_\infty$ (see also Refs.~\cite{Ruffini:1972uh,Cardoso:2003cn}) while the rate of convergence of the series over the partial energies $E_\ell$ which defines it decreases. In other terms, i.e., from a physical point of view, we can observe that for $v_\infty = 0$, the $\ell = 1$ mode radiates the largest amount of energy ($83.20\%$), and that summing over the first five modes, we reach $99.99\%$ of the total electromagnetic energy radiated; on the other hand, for $v_\infty = 0.99$, the $\ell = 1$ mode is responsible for only $16.54\%$ of the total electromagnetic energy radiated while the sum over the first five modes represents only $63.41\%$ of this energy.

\begingroup
\squeezetable
\begin{table}[H]
\caption{\label{tab:table1} The total energy $E$ radiated by the charged particle is considered for the values $v_\infty =0, 0.75, 0.90$ and $0.99$ of its velocity at infinity. The percentage of energy radiated in the $\ell=1$ mode and in the first five modes is also considered. Here, we have taken $2M=1$.}
\centering
\begin{ruledtabular}
\resizebox{\columnwidth}{!}{%
\begin{tabular}{lccc}
     $v_\infty\, (\gamma)$  & $(1/q)^2 E $  & $(1/q)^2 E_{1} $ &$(1/q)^2 \sum_{\ell= 1}^{5} E_{\ell}$  \tabularnewline

    \hline
    $ 0\, (1)   $ & $ 3.4049 \times 10^{-3}$ & $83.20\%$ & $99.99 \% $  \tabularnewline
    $ 0.75\, (1.51)$ & $ 1.0181 \times 10^{-2}$ & $69.20\%$ & $99.83 \% $  \tabularnewline
    $ 0.90 \, (2.29)$ & $ 2.3251 \times 10^{-2}$ & $49.45\%$ & $97.68 \% $  \tabularnewline
    $ 0.99\, (7.09)$ & $ 1.0726 \times 10^{-1}$ & $16.54\%$ & $63.41 \% $  \tabularnewline
\end{tabular}
}
\end{ruledtabular}
\end{table}
\endgroup

In Fig.~\ref{dEdw_Exact_vs_CAM_v_0_075_090_099}, we have compared the electromagnetic energy spectrum $dE/d\omega$ given by Eq.~(\ref{dEdw_bis}) with its CAM representation (\ref{dEdw_CAM_2})-(\ref{dEdw_CAM_PR}). This permits us to emphasize the respective role of the background integrals (\ref{dEdw_background_integral_Re}) and (\ref{dEdw_background_integral_Im}) and of the Regge pole sum (\ref{dEdw_CAM_PR}). In particular, we can observe that, for very low frequencies, in order to match the exact energy spectrum, it is necessary to take into account these two background integrals and to consider the first two Regge poles in the Regge pole sum. Out of this frequency regime, the exact energy spectrum can be perfectly described by only considering the background integral along the real axis and a single Regge pole in the Regge pole sum. Here, it is worth pointing out that the Regge pole approximation cannot be used to resum the total electromagnetic energy spectrum because the CAM representation is dominated by the background integrals. However, we can observe in Fig.~\ref{dEdw_Exact_vs_CAM_PR_Int_axe_Re_v_0_090_099} that it is the Regge pole approximation which explains the oscillations appearing in the electromagnetic energy spectrum. Due to the connection existing between the Regge modes and the (weakly damped) QNMs of the Schwarzschild BH \cite{Decanini:2002ha,Decanini:2009mu}, we can also associate these oscillations with the quasinormal frequencies of the BH.

\newpage

\section{Conclusion}
\label{Conc}

In this paper, we have revisited the problem of the electromagnetic radiation generated by a charged particle falling radially into a Schwarzschild BH. We have obtained a series of results which highlight the benefits of working within the CAM framework and strengthen our opinion concerning the interest of the Regge pole approach for describing radiation from BHs because they are fairly close to those previously reported in Ref~\cite{Folacci:2018sef} where we discussed an analogous problem in the context of gravitational radiation.

We have described the electromagnetic radiation by the Maxwell scalar $\phi_2$ and we have extracted from its multipole expansion~\eqref{phi2_ExpressionDef_reg} the Fourier transform of a sum over the Regge poles of the BH $\mathcal{S}$-matrix involving, in addition, the excitation factors of the Regge modes. It constitutes an approximation of $\phi_2$ which can be evaluated numerically from the Regge trajectories associated with the Regge poles and their residues. In fact, we have constructed two different Regge pole approximations of $\phi_2$: the first one, which has been obtained from the Poisson summation formula, is given by Eq.~\eqref{CAM_phi2_ExpressionDef_P_RP} and provides very good results (even impressive results for relativistic particles) for observation directions in a large angular sector around the particle trajectory; the second one, which has been derived by using the Sommerfeld-Watson transformation, is given by Eq.~\eqref{CAM_phi2_ExpressionDef_SW_RP} and is helpful in a large angular sector around the direction opposite to the particle trajectory. More precisely, it should be noted that, in general, these two Regge pole approximations can reproduce with very good agreement the quasinormal ringdown (it is worth pointing out that, in contrast to the QNM description of the ringdown, the Regge pole description does not require a starting time) as well as with rather good agreement the tail of the signal and that the first approximation even describes the pre-ringdown phase. All our results have been achieved by taking into account only one Regge pole. To understand the interest of this fact, it is important to recall that the partial wave expansion defining $\phi_2$ is a slowly convergent series, especially in the case of a particle projected with a relativistic or an ultra-relativistic velocity into the BH; its Regge pole approximations are efficient resumations which permit us, in addition, to extract the physical information it encodes. It is interesting to recall that, for the analogous problem in the context of gravitational radiation \cite{Folacci:2018sef}, we have obtained rather similar results for the Weyl scalar $\Psi_4$ but that, in this case, taking into account additional Regge poles sometimes improves the Regge pole approximations. It should be finally noted that we have also considered the electromagnetic energy spectrum $dE/d\omega$ (a topic we did not touch on in Ref~\cite{Folacci:2018sef}) and, by using the Poisson summation formula, we have constructed from its multipole expansion~\eqref{dEdw_bis} its CAM representation given by Eqs.~(\ref{dEdw_CAM_2})--(\ref{dEdw_CAM_PR}). Unfortunately, here the full CAM representation is necessary to describe the whole electromagnetic energy spectrum but the corresponding Regge pole approximation (\ref{dEdw_CAM_PR}) is however helpful to understand its oscillations and associate them with QNMs.

In future works, we would like to go beyond the relatively simple problems examined in the present paper and in Ref.~\cite{Folacci:2018sef} by revisiting, using CAM techniques, the problem of the radiation generated by a particle with an arbitrary orbital angular momentum plunging into a Schwarzschild or a Kerr BH. It would be in addition interesting to extract asymptotic expressions from the background integral contributions appearing in the various CAM representations in order to improve the physical interpretation of the results. We would also like to go beyond the case of BHs by considering that of neutron stars and
white dwarfs. In this context, the recent CAM analysis of scattering by compact objects \cite{OuldElHadj:2019kji} could be a natural starting point.

%
%
%

%

\bibliography{RP_RT_EM}

\end{document}